\documentclass[%
onecolumn,   
 notitlepage,
 11pt,
nofootinbib,
 amsmath,amssymb,
 aps,
]{revtex4-2}

\usepackage{graphicx}
\usepackage{dcolumn}
\usepackage{bm}
\usepackage[dvipsnames]{xcolor}
\usepackage[colorlinks=true,
            citecolor=ForestGreen,
            linkcolor=ForestGreen,
            urlcolor=ForestGreen]{hyperref}
\usepackage{setspace}
\usepackage{amsmath}

\begin{document}

\newcommand{\ms}[1]{{\color{black} {#1}}}

\def\IB#1{\boldsymbol{#1}} 
\def\re#1{\Re_{#1}} 
\def\Wi{\text{Wi}\,} 
\def\bten#1{\IB{\mathsf{#1}}}

\renewcommand{\thefootnote}{\fnsymbol{footnote}}




	
	\title{
		Tuning Cross-stream Lift in Viscoelastic Shear: Distinct Hydrodynamic Signatures of Force-bearing and Force-free Mechanisms
	}
	
\author{Soumyodeep Chowdhury}
\author{Kushagra Tiwari}
\author{Jitendra Dhakar}
\author{Akash Choudhary}
\email{achoudhary@iitk.ac.in}
	%
	\affiliation{Department of Chemical Engineering, Indian Institute of Technology Kanpur, India}
	%
	\begin{abstract}
		We investigate the lift and drag corrections acting on a particle suspended in a planar viscoelastic shear flow when the particle is tuned to translate relative to the flow by an external mechanism. 
		A cross-stream lift force arises when particle is driven in streamwise direction; we find that the nature of the driving mechanism dictates the lift direction: force-bearing mechanisms (such as gravity acting on non-neutrally buoyant particles) and force-free mechanisms (such as electrophoresis) generate lift forces of opposite sign.
		By explicitly deriving the first-order fields and stresses, we demonstrate that this reversal originates from distinct hydrodynamic disturbances induced by each mechanism, which produce qualitatively different polymeric stress distributions. This analytical result is further verified through an independent derivation using the reciprocal theorem.
		Further, we find that driving the particle in the gradient direction gives rise to a streamwise drag correction that is of the same sign for both mechanisms.
		Beyond microfluidic particle manipulation, these results have broader implications 
		for understanding the locomotion of microswimmers in viscoelastic shear flows, 
		where distinct force-free propulsion mechanisms are expected to generate unique force and torque modifications.
	\end{abstract}
	\maketitle
	
	
	\section{Introduction}
	Particles suspended in creeping flows of viscoelastic fluid can migrate across streamlines due to the non-linear nature of polymeric stresses.
	In pressure-driven flows, both early experimental observations \citep{karnis1966particle}, theoretical predictions \citep{ho1976migration,brunn1976slow} and simulations \citep{huang1997direct} established that particles migrate toward the channel centreline. 
	Further \citet{leshansky2007tunable} showed that, in microfluidic devices, desired particle focusing can be achieved solely through rheological control of the viscoelastic carrier solution.
	These examples of viscoelastic migration in channel flows share a common feature: the background flow possesses \ms{curvature in the velocity profile}, which generates hoop stresses arising from gradients in the first normal stress difference (largest near the walls and diminishing toward the centreline) that drive cross-stream particle drift.
	In contrast, simple shear flow cannot passively drive this migration. However, buoyancy-driven relative motion between the particles and mean flow can yield inhomogeneities in sedimenting suspensions in weakly viscoelastic fluids, manifesting as particle clumping and aggregation \citep{Phillips2010,Vishnampet2012}. 
	Subsequently, using perturbation expansion, \citet{einarsson2017spherical} revisited this problem \citep{brunn1977errata} for an Oldroyd-B fluid until the second order Weissenberg number.
	They found that applying a flow-parallel gravitational field yields cross-stream migration of buoyant particles: a leading particle migrates towards regions of high velocity, whereas a lagging particle migrates towards lower velocity regions. 
	\citet{Zhang2020} performed simulations for Oldroyd-B sheared flow to quantify the lift force on a sphere across weak to moderate Weissenberg numbers. 
	They demonstrated that this lift force arises from an imbalance in polymer stretch surrounding the particle, driven by the asymmetry in the local relative velocity.
	While these results suggest that migration can be systematically controlled via buoyancy tuning, the required vertical experimental setups often impose practical constraints.
	Consequently, recent microfluidic studies have explored the use of electric fields as an alternative approach to tuning particle migration \citep{lu2017particle}.

	\begin{figure}
		\centerline{
			\includegraphics[width=0.95 \linewidth]{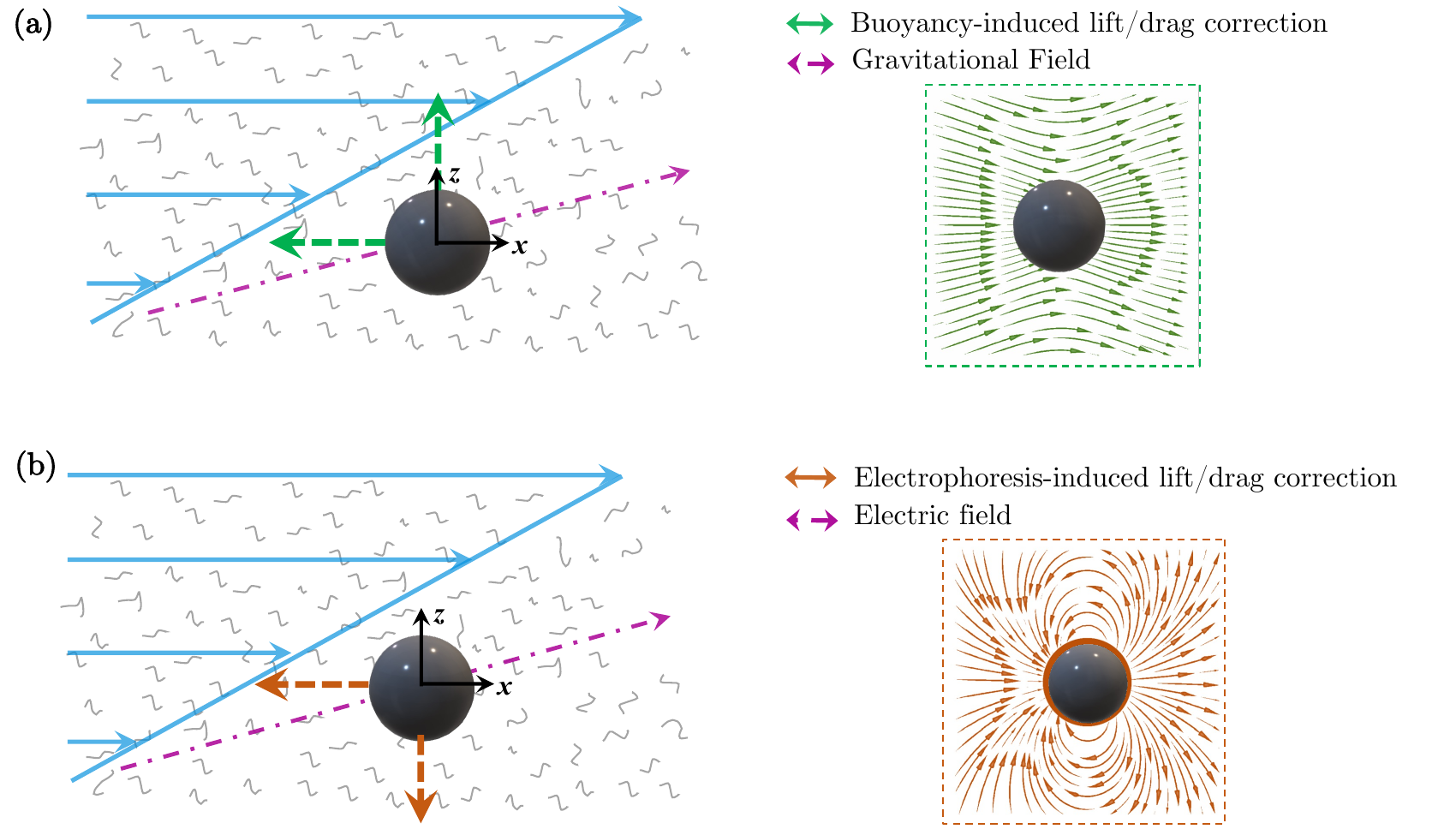}
		}
		\caption{ 
        \ms{Schematic depicts a particle suspended in viscoelastic shear flow subjected to (a) gravitational field $\IB{\hat{g}}_\infty$  (buoyancy-tuned mechanism) and (b) electric field $\IB{\hat{E}}_\infty$ (electrophoresis-tuned mechanism).
			The origin of the coordinate system is at the center of the particle that is made to move relative to the flow using external field.
            The $x$ (and $z$) components of the relative velocity result in lift (and drag) contribution.
			Figures on the right of (a,b) depict the hydrodynamic signature imparted by the addition of two kinds of external fields.}
            }
		\label{fig:schematic}
	\end{figure}

	Experiments in a pressure-driven viscoelastic flow by \citet{Li2018} showed that the cross-stream migration of electrophoretic particles follows the trend predicted by the buoyancy-driven lift mechanism of \citet{einarsson2017spherical}.
	However, these experiments were conducted at PEO concentrations exceeding the overlap concentration, placing the solution in the semi-dilute regime where \ms{polymer chains entangle} to exhibit shear-thinning effects that would deviate from weakly elastic perturbation theory. 
	\ms{Their subsequent experimental work in strictly dilute PEO concentrations reported that electrokinetically-enhanced lift and modeled this effect via the reciprocal theorem framework for weakly viscoelastic flows \citep{Choudhary20}.
    However, while the reciprocal theorem serves as a powerful mathematical tool to compute lift at first order (without having to explicitly evaluate its hydrodynamics), it relies on an integral formulation wherein all weakly non-linear stress contributions are subsumed within a global volume integral.
    This mathematical convenience obscures the details of the first-order hydrodynamics and the spatial distribution of polymeric stresses, which leaves the physical mechanism unresolved \citep[p. 795]{ho1976migration}. 
    Consequently, a conceptual gap persists in the literature regarding why migration differs between buoyancy and electrophoretic scenarios, leading to a conflation of the two mechanisms.}
	\ms{A recent hybrid (course-grained molecular dynamics and Lattice-Boltzmann) computational} study by \citet{Ma2026} finds that lift scales linearly with relative tuning velocity; however, their \ms{intensive simulations drive the particle via a direct body force (that would yield a stokeslet signature), whereas their experiments rely on electrophoresis, a force-free mechanism (characterized by a source-dipole signature). 
    Because these mechanisms induce distinct hydrodynamic disturbances and polymeric stresses, these underlying fundamental differences can get masked by attempts at unified scaling approaches based solely on relative tuning velocity.}
    Furthermore, subsequent experiments from \citet{Serhatlioglu2020} and \citet{Li2023} reveal that migration direction switches signs for polymeric concentration above and below the overlap concentration. Interestingly, they observed that when electrophoretic velocity is increased to match the shear-rate velocity scale $\dot{\gamma}a$ (\textit{i.e.,} De$\sim$Wi), this onset is delayed.
	
	\ms{The goal of the current work is to distinguish between force-bearing and force-free mechanisms in the context of external-field-tuned particle migration, focusing specifically on the weakly elastic regime. For this, we employ the second-order fluid model because it yields an accurate asymptotic approximation for the majority of weakly elastic steady flows \cite{bird1986dynamics}.}
	Fig. \ref{fig:schematic} schematically illustrates the two distinct cases studied within a single theoretical framework that captures the effects of weak viscoelasticity via perturbation theory. 
  \ms{
  The general governing equations and specific boundary conditions are outlined in section II.
  In section III, we explicitly calculate the first-order stress fields by applying the framework by \citet{Peery1966} for inhomogeneous Stokes equations. By resolving the local hydrodynamics, the analytical results explain how different tuning mechanisms generate distinct polymer stretching around the particle. The distinction in the external mechanism is found to be crucial: the resulting elastic lift and its direction depend fundamentally on the nature of the flow disturbance rather than the magnitude of relative velocity alone.  
  Additionally, we also derive the first order correction to drag in flow direction, and find that first order force corrections are symmetric (\textit{i.e.,} equal lift and drag corrections) for electrophoretic-tuning because its leading order hydrodynamic signature is irrotational. Whereas, since a stokeslet exhibits local vorticity, the buoyancy-induced tuning yields an asymmetric force correction. 
  Section IV details the physical interpretation of results, comparison with literature, and future outlook.
  \\.
  %
  }

	\section{Problem formulation}
	Figure \ref{fig:schematic} shows a spherical particle suspended in uniform unbounded planar shear flow of a second-order fluid.
	We assume this weakly viscoelastic flow to be in the Stokes regime \textit{i.e.,} $Re\ll1$ with $Re=\frac{\rho \dot{\gamma} a^2 }{\mu}$ with  $\dot{\gamma}$, $a$, $\rho$ and $\mu$ represents shear rate, particle's radius, fluid density and viscosity,  respectively. 
	The coordinate system is assigned such that the streamwise direction, cross-stream direction, and vorticity direction correspond to the $x-$, $z-$, and $y-$axes, respectively. 
	The particle's motion is described by its translational velocity $ \IB{V}_s$ and angular velocity $\IB{\Omega}_s$ and the analysis is carried out in the particle’s reference frame of translation.
	The spatial coordinates, velocity, and pressure are non-dimensionalized using $a$, $\dot{\gamma} a$, and $\mu \dot{\gamma}$, respectively.
    
	\ms{In this work, we employ the resistance formulation, and calculate the response of a particle when it is subjected to an external mechanism that imparts a velocity ($\IB{V}_r$) relative to the flow}: $\IB{V}_s = \left(1 + \frac{\nabla^2}{6}\right)\IB{v}^\infty \vert_{\IB{r}=0} + \IB{V}_r $, where the first term comes from Faxén's law \citep{KimKarrila13}, which denotes the particle velocity imparted by the ambient flow in the absence of external mechanisms. 
	We have the background shear flow as $(\alpha+\beta z) \IB{\hat{e}}_x$, where $\alpha$ is the uniform flow component and $\beta$ represents the strength of shear. {Although $\beta = 1$ for uniform shear, we retain it in the formulation to clearly represent the shear contribution in the forthcoming expressions}.
	\ms{Note that in resistance formulation, the prescribed velocity, by definition, incorporates no higher-order corrections. Consequently, this approach is strictly equivalent to the mobility formulation at the $O(\text{Wi})$ approximation \cite{ho1976migration,einarsson2017spherical}, which is within the scope of the present study.}
    We investigate how and why the lift and torque corrections differ when the imposed relative velocity is tuned by buoyancy as opposed to electrophoresis.
	Below we describe the general system of equations for fluid motion that apply for both cases. We will study each case individually after defining the  governing equations:
	\begin{align}\label{eq:model-eq}
		\nabla^2 \IB{v} - \nabla p = -\Wi  \nabla \cdot &\bten{\Pi}, \quad \nabla \cdot \IB{v} = 0, 
		\\
		\IB{v} = \IB{\Omega}_s \times \IB{r} + \mathcal{E} \nabla \phi - \IB{V}_\infty \text{ at } r = 1 & \quad \text{ and } \quad
		\IB{v} \to \ms{\IB{0}} \text{ as } r \to \infty, \nonumber
	\end{align}
	where, $\IB{v}$ and $p$ denote the disturbed flow and pressure fields respectively. 
	$\mathcal{E}=\epsilon \zeta E_\infty/(\mu \dot{\gamma} a) $ denotes the dimensionless electrophoretic mobility, where $\epsilon, \, \zeta, \, E_\infty $ are permittivity, zeta potential and magnitude of electric field, respectively. 
	The potential field $\phi$ around a typical polystyrene particle is such that $\nabla \phi =-\frac{3}{2}(\bten{I}-\IB{nn})\cdot  \IB{\hat{E}}_\infty $, where $\IB{\hat{E}}_\infty$ is the unit vector of its direction \citep{anderson1989colloid}.
	The undisturbed flow profile in the particle frame of reference is given by
	$\IB{V}_\infty = (\alpha +  \beta z) \IB{\hat{e}}_x - \IB{V}_s$.
    \ms{The rigid body motion of sphere $(\IB{V}_s$ and $\IB{\Omega_s})$ will be evaluated via force and torque balance equations at the leading order.}
	\ms{The disturbance polymeric stress is denoted by $\bten{\Pi}$} that captures weak elastic effects.
	The shear-based Weissenberg number ($\Wi$) denotes the ratio of viscoelastic relaxation time scale $({\Psi_1+\Psi_2})/{\mu}$ to the time scale of shear $\dot{\gamma}^{-1}$, where $\Psi_i$ denotes the normal stress coefficients.
	The steady non-Newtonian stress contribution can be expressed as sum of co-rotational and quadratic contributions \ms{(detailed in Appendix A): $\bten{\Pi} = \bten{\Pi}^{C} + \bten{\Pi}^{Q}$}, where 
	\begin{align}\label{eq:PI}
		\bten{\Pi}^{C} &= 
		2 \delta \Big(
		\nabla \cdot (\IB{v}\,\bten{e})
		+ \nabla \cdot (\IB{V}_\infty\,\bten{e})
		+ \nabla \cdot (\IB{v}\,\bten{e}_\infty)
		+ \bten{\omega} \cdot \bten{e}
		+ \bten{e} \cdot \bten{\omega}^{\mathsf{T}}
		\\
		&\qquad\qquad\qquad \qquad \quad
		+ \bten{\omega}_\infty \cdot \bten{e}
		+ \bten{e} \cdot \bten{\omega}_\infty^{\mathsf{T}}
		+ \bten{\omega} \cdot \bten{e}_\infty
		+ \bten{e}_\infty \cdot \bten{\omega}^{\mathsf{T}}
		\Big),\nonumber
		\\
		\bten{\Pi}^{Q} &= 
		4 (1+\delta)\  \Big(
		\bten{e} \cdot \bten{e} +\bten{e}_\infty \cdot \bten{e}+\bten{e} \cdot \bten{e}_\infty
		\Big).
	\end{align}
	Here, $\delta$ denotes the viscometric coefficient $-\Psi_1/2(\Psi_1+\Psi_2)$ (mostly between -0.5 and -0.6),
	\(\bten{e}\) and \(\bten{e}_\infty\) denote the symmetric rate-of-strain tensors associated with the disturbance and undisturbed flow fields, respectively, while \(\bten{\omega}\) and \(\bten{\omega}_\infty\) are the corresponding antisymmetric vorticity tensors. 
    \ms{The co-rotational component captures stresses generated by the rotation and convection of the polymeric microstructure, whereas the quadratic component ($\boldsymbol{\Pi}_Q$) represents stresses generated by symmetric deformation of the microstructure \cite{bird1986dynamics}.}

\ms{\subsection{Perturbation expansion}}
    
	Here we restrict our attention to the weak viscoelastic behavior of the fluid and employ an asymptotic expansion in Wi$\ll 1$ regime to expand velocity and pressure fields in Wi: $\IB{v} = \IB{v}^{(0)} + \Wi\, \IB{v}^{(1)} + \cdots, \; \; p = p^{(0)} + \Wi\, p^{(1)} + \cdots$. 
	Consequently, Eq.(\ref{eq:model-eq}) breaks down at each order as follows:
	\begin{align}\label{eq:model_eq_zero}
		& \nabla^{2}\IB{v}^{(0)} - \nabla p^{(0)} = 0,
		\qquad 
		\nabla \cdot \IB{v}^{(0)} = 0, 
		\\
		\IB{v}^{(0)} &= \IB{\Omega}_s \times \IB{r} 
		+ \mathcal{E} \nabla \phi
		- \IB{V}_\infty
		\; \text{at } r=1, 
		\quad \text{and} \quad 
		\IB{v}^{(0)} \to \ms{\IB{0}}
		\; \text{as } r\to\infty. \nonumber
	\end{align}
	\begin{align}\label{eq:model_eq_first}
		& \nabla^{2}\IB{v}^{(1)} - \nabla p^{(1)}
		= -\,\nabla \cdot \bten{\Pi}^{(0)},
		\qquad
		\nabla \cdot \IB{v}^{(1)} = 0, 
		\\
		\IB{v}^{(1)} &= 0
		\; \text{at } r=1, \quad \text{and} \quad \IB{v}^{(1)} \to 0
		\; \text{as } r\to\infty.\nonumber
	\end{align}
	This perturbation expansion in shear-based Weissenberg number assumes that the \ms{relative} velocity is of the same order as $\dot{\gamma}a$, a regime analogous to that of \citet{einarsson2017spherical}.
	The zeroth-order flow field can be derived analytically using Lamb's general solution or vector harmonics \citep{guazzelli2011physical}.
	We present the solution as a superposition of the effect of terms in the boundary condition as:
	\begin{align}\label{eq:V0_decomposition}
		\IB{v}^{(0)} &= 
		\frac{(\IB{\Omega}_s-  \IB{\Omega}_\infty ) \times \IB{r}}{r^{3}}
		+
		\frac{3}{4}\,
		( \IB{V}_s - \alpha \IB{\hat{e}}_x )
		\cdot
		\left(
		\frac{\bten{I}}{r} + \frac{\IB{rr}}{r^{3}} 
		+
		\frac{\bten{I}}{3r^3}- \frac{\IB{rr}}{r^{5}}
		\right)
		\nonumber
		\\
		&
		-
		\beta \frac{ \bten{e}_\infty \cdot \IB{r} }{ r^{5} } 
		-\frac{5\beta}{2}
		\big( \bten{e}_\infty : \IB{r}\IB{r} \big) 
		\IB{r}
		\left( \frac{1}{r^{5}} - \frac{1}{r^{7}} \right)
		-
		\frac{3}{4} \mathcal{E} \IB{\hat{E}}_\infty \cdot
		\left(
		\frac{\bten{I}}{r}
		+
		\frac{\IB{r}\IB{r}}{r^{3}}
		+
		\frac{\bten{I}}{r^{3}}
		-
		\frac{3\IB{r}\IB{r}}{r^{5}}
		\right)
		. \nonumber
		\\
		p^{(0)} &= \frac{3}{2} \frac{( \IB{V}_s - \alpha \IB{\hat{e}}_x )\cdot \IB{r}}{r^3} 
		- 5 \frac{\bten{e}_\infty : \IB{rr}}{r^5} 
		- \frac{3}{2} \frac{\mathcal{E} \IB{\hat{E}}_\infty \cdot \IB{r}}{r^3} .
	\end{align}
	Here $\IB{\Omega}_\infty $ denotes the angular velocity associated with the antisymmetric part of the imposed far-field flow.
	We now use the force and torque balance conditions at zeroth order to determine $\IB{V}_s$ and $\IB{\Omega}_s$ (note that force and torque are scaled using $\mu \dot{\gamma} a^2$ and $\mu \dot{\gamma} a^3$).
	We perform this balance separately for buoyancy and electrophoretic tuning.
	
\ms{\subsection{Force and Torque balance at $O(1)$}}

\ms{The total translational and angular velocities in Eq. (\ref{eq:V0_decomposition}), which have thus far remained unknown, are subsequently derived using the $O(1)$ force and torque balance conditions.}
	For buoyant particle, we impose $\mathcal{E}=0$ and obtain the following expression balancing the scaled buoyant force ($\IB{F}_{\text{buoyancy}} = - \IB{F}_{\text{drag}}$):
	\begin{equation}\label{eq:zero-order_particle_vel_buoyancy}
		\left( \frac{4\pi a g \Delta\rho}{3 \mu \dot{\gamma}} \right)  \IB{\hat{g}}_\infty = - \int_{S_p} \IB{n} \cdot (-p^{(0)}\bten{I} + 2 \bten{e}^{(0)}) dS = - 6\pi (\alpha \IB{\hat{e}}_x - \IB{V}_s)
	\end{equation}
	where $\IB{\hat{g}}_\infty$ represents unit vector in the direction of applied field.
	Noting that $\IB{V}_s = \alpha \IB{\hat{e}}_x + \IB{V}_r$ for shear flow and defining the dimensionless buoyancy number, we get:
	\begin{equation}\label{eq:zero-order_particle_translation-B}
		\IB{V}_r = \mathcal{B} \, \IB{\hat{g}}_\infty, \text{ where } \mathcal{B} = {2 a \Delta \rho g}/({9 \mu \dot{\gamma}}).
	\end{equation}
	For electrophoretic particle, the force-free condition gives: $0=-6 \pi (\alpha \IB{\hat{e}}_x - \IB{V}_s) - 6 \pi \mathcal{E} \IB{\hat{E}}_\infty$, which yields:
	\begin{equation}\label{eq:zero-order_particle_translation-EP}
		\IB{V}_r = \mathcal{E} \IB{\hat{E}}_\infty, \text{ where } \mathcal{E}=\epsilon \zeta E_\infty/(\mu \dot{\gamma} a)
	\end{equation}
	is the electrophoretic mobility as described in Eq.(\ref{eq:model-eq}).
	Finally, we note that under the torque-free constraint, we get:
	\begin{equation}
		\IB{\Omega}_s= \IB{\Omega}_\infty = \tfrac12\nabla\times\IB{V}_\infty = \tfrac{\beta}{2} \IB{\hat{e}}_y.
		\label{eq:zero-order_particle_rot}
	\end{equation}
	for both tuning mechanisms. 
	The zeroth order fields for buoyant particle case can be completed by substituting Eq.(\ref{eq:zero-order_particle_translation-B}, \ref{eq:zero-order_particle_rot}) into Eq.(\ref{eq:V0_decomposition}) for $\mathcal{E}=0$, whereas fields for electrophoretic particle can be obtained by substituting Eq.(\ref{eq:zero-order_particle_translation-EP}, \ref{eq:zero-order_particle_rot}) into Eq.(\ref{eq:V0_decomposition}).



	\section{First order contributions to lift and drag}
	
	\subsection{Buoyant particle in simple shear}
	
	Since the aim is to find solution to the inhomogeneous Eq.(\ref{eq:model_eq_first}), we split the the stress on the right hand side into quadratic and co-rotational components and evaluate them separately and follow approach similar to that outlined in \citet{Peery1966,KOCH200687}. As an illustration of its utility, we outline the solution procedure below and  all the steps and coefficients are detailed in Appendix B.
	Breaking down $\bten{\Pi}^{(0)}$ into quadratic and co-rotational components, the field equation governing the quadratic contribution is:
	\begin{equation}
		\label{eq:quad GE}
		\nabla^2 \IB{v}^{(1Q)} - \nabla p^{(1Q)} =- \nabla \cdot {\bten{\Pi}^{(0Q)}}, \quad \nabla \cdot \IB{v}^{(1Q)}=0
	\end{equation} 
	where $\bten{\Pi}^{(0Q)}$ is evaluated in terms of $\IB{v}^{(0)}$.
	Next, by taking divergence on both sides we obtain an equation governing pressure component:
	\begin{align}
		\label{eq:p1Q-GE}
		&\nabla^2 p^{(1Q)} = \nabla\nabla:\bten{\Pi}^{(0Q)}, \text{ where }
		\\
		\label{eq:p1Q-rhs}
		&\nabla\nabla:\bten{\Pi}^{(0Q)}
		=
		4(1+\delta)\left[
		f_1\IB{V}_r\cdot\IB{V}_r
		+ f_2(\IB{r}\cdot\IB{V}_r)(\IB{r}\cdot\IB{V}_r)
		+ f_3(\bten{e}_\infty:\IB{r}\IB{V}_r)
		\right.
		\\
		&
		\qquad
		\left.+ f_4(\bten{e}_\infty:\IB{r}\IB{r})(\IB{r}\cdot\IB{V}_r)
		+ f_5(\bten{e}_\infty\cdot\IB{r})\cdot(\bten{e}_\infty\cdot\IB{r})
		+ f_6(\bten{e}_\infty:\IB{r}\IB{r})^2
		+ f_7\bten{e}_\infty:\bten{e}_\infty
		\right].
		\nonumber
	\end{align}
	Here, $f_i(r)$ are known coefficients detailed in Appendix B. Based on the structure of these seven non-homogeneous terms, an ansatz for $p^{(1Q)}$ is constructed with unknown coefficients $g_i(r)$. Substituting this ansatz into Eq.~(\ref{eq:p1Q-GE}) yields seven second-order ordinary differential equations in $g_i(r)$, which are solved to obtain the particular and homogeneous solutions for $p^{(1Q)}$.
	\begin{align}
		&{p}^{(1Q)}
		=
		4(1+\delta)\left[
		g_1\IB{V}_r\cdot\IB{V}_r
		+ g_2(\IB{r}\cdot\IB{V}_r)^2
		+ g_3(\bten{e}_\infty:\IB{r}\IB{V}_r)
		+ g_4(\bten{e}_\infty:\IB{r}\IB{r})(\IB{r}\cdot\IB{V}_r)
		\nonumber
		\right.
		\\
		&\qquad
		\left.+ g_5(\bten{e}_\infty\cdot\IB{r})\cdot(\bten{e}_\infty\cdot\IB{r})
		+ g_6(\bten{e}_\infty:\IB{r}\IB{r})^2
		+ g_7\bten{e}_\infty:\bten{e}_\infty
		\right],
	\end{align}
	where the coefficients $g_i(r)$ comprise homogeneous and particular components. The integration constants of the homogeneous component remain undetermined at this stage and are resolved concurrently with the velocity field upon enforcement of the boundary conditions.
	Substitution of the quadratic pressure field into Eq.~\eqref{eq:quad GE} yields $\nabla^2 \IB{v}^{(1Q)} = \nabla p^{(1Q)}-\nabla\cdot\bten{\Pi}^{(0Q)}$, 
    \ms{where the right-hand side has a tensorial structure with coefficients $h_i (r)$, which comprise of: $g_i(r)$, its derivatives, and fully known functions from the expansion of $\nabla \cdot \bten{\Pi}^{(0Q)}$.
    This facilitates an ansatz for $\IB{v}^{(1Q)}$ with unknown coefficients in $j_i (r)$}.
	Following the similar procedure as pressure field, we substitute the ansatz in $\nabla^2 \IB{v}^{(1Q)}$. \ms{This results in a system of straightforward ordinary differential equations in $j_i(r)$}, whose solution yields the particular and homogeneous solutions, where the latter is determined using the application of boundary condition Eq. (\ref{eq:model_eq_first}) and continuity equation.  We obtain the quadratic velocity correction as:
	\begin{align}\label{eq:v1Q_ansatz}
		&\IB{v}^{(1Q)}
		= 4(1+\delta)[
		j_1\IB{r}(\IB{V}_r\!\cdot\!\IB{V}_r)
		+ j_2\IB{V}_r(\IB{V}_r\!\cdot\!\IB{r})
		+ j_3\IB{r}(\IB{V}_r\!\cdot\!\IB{r})^{2}
		+ j_4(\bten{e}_{\infty}\!\cdot\!\IB{V}_r)
		\\
		&
		+ j_5(\bten{e}_{\infty}\!\cdot\!\IB{r})(\IB{V}_r\!\cdot\!\IB{r})
		+ j_6\IB{V}_r(\bten{e}_{\infty}\!:\!\IB{r}\IB{r})
		+ j_7\IB{r}(\bten{e}_{\infty}\!:\!\IB{r}\IB{V}_r)
		+ j_8\IB{r}(\IB{V}_r\!\cdot\!\IB{r})(\bten{e}_{\infty}\!:\!\IB{r}\IB{r})
		\nonumber
		\\
		&
		+ j_9\bten{e}_{\infty}\!\cdot(\bten{e}_{\infty}\!\cdot\!\IB{r})
		+ j_{10}\IB{r}(\bten{e}_{\infty}\!\cdot\!\IB{r})^{2}
		+ j_{11}(\bten{e}_{\infty}\!\cdot\!\IB{r})(\bten{e}_{\infty}\!:\!\IB{r}\IB{r})
		+ j_{12}\IB{r}(\bten{e}_{\infty}\!:\!\IB{r}\IB{r})^{2}
		+ j_{13}\IB{r}(\bten{e}_{\infty}\!:\!\bten{e}_{\infty})
		].\nonumber
	\end{align}

	The co-rotational component is governed by:
	\begin{equation}
		\label{eq:C GE}
		\nabla^2 \IB{v}^{(1C)} - \nabla p^{(1C)} =- \nabla \cdot {\bten{\Pi}^{(0C)}}, \quad \nabla \cdot \IB{v}^{(1C)}=0
	\end{equation} 
	Following the similar approach as above, we obtain the pressure contribution as:
	\begin{align}
		&p^{(1C)} 
		= 
		\delta \left[
		k_{1}(\IB{r}\cdot\IB{V}_r)^2
		+ k_{2}(\bten{e}_\infty:\IB{r}\IB{V}_r)
		+ k_{3}(\bten{e}_\infty:\IB{r}\IB{r})(\IB{r}\cdot\IB{V}_r)
		+ k_{4}(\bten{e}_\infty\cdot\IB{r})\cdot(\bten{e}_\infty\cdot\IB{r}) \nonumber
		\right.
		\\
		&\qquad
		\left. + k_{5}(\bten{e}_\infty:\IB{r}\IB{r})^2
		+ k_{6}(\bten{\omega}_\infty:\IB{r}\IB{V}_r)
		+ k_{7}(\bten{e}_\infty\cdot\IB{r})\cdot(\bten{\omega}_\infty\cdot\IB{r})
		+ k_{8}(\bten{e}_\infty:\bten{e}_\infty)
		\right]
	\end{align}
	Here $k_i(r)$ are the known coefficients where $k_i(r)$ $(i=1,\ldots,8)$ are given in Appendix B.
	The velocity contribution $\IB{v}^{(1C)}$ vanishes \citep{KOCH200687}, as the divergence of the co-rotational stress reduces to the gradient of a scalar, thereby only modifying the pressure field under homogeneous boundary conditions for $\IB{v}^{(1C)}$ (see Supplementary material for this derivation).


	\subsection{Electrophoretic particle in simple shear}
	For this force-free case, we follow the similar approach as above i.e. we again solve the inhomogeneous Stokes equation Eq.(\ref{eq:model_eq_first}) to find the ${O}(\text{Wi})$ velocity $\IB{v}^{(1Q)}$ and pressure fields $p^{(1Q)}$, $p^{(1C)}$ respectively.
	The details being analogous, only the final expressions are reported below.
	\begin{align}
		&p^{(1Q)}
		=
		4(1+\delta)\left[
		\Tilde{g}_1\IB{V}_r\cdot\IB{V}_r
		+ \Tilde{g}_2(\IB{V}_r\cdot\IB{r})^2
		+\Tilde{g}_3(\bten{e}_\infty:\IB{V}_r\IB{r})
		+ \Tilde{g}_4(\bten{e}_\infty:\IB{r}\IB{r})(\IB{V}_r\cdot\IB{r})
		\nonumber
		\right.
		\\
		&\qquad
		\left.+\Tilde{g}_5(\bten{e}_\infty\cdot\IB{r})\cdot(\bten{e}_\infty\cdot\IB{r})
		+ \Tilde{g}_6(\bten{e}_\infty:\IB{r}\IB{r})^2
		+ \Tilde{g}_7\bten{e}_\infty:\bten{e}_\infty
		\right],
	\end{align}
	\begin{align}
		&p^{(1C)}
		=
		\delta\left[
		\Tilde{k}_1\IB{V}_r\cdot\IB{V}_r
		+ \Tilde{k}_2(\IB{V}_r\cdot\IB{r})^2
		+\Tilde{k}_3(\bten{e}_\infty:\IB{V}_r\IB{r})
		+ \Tilde{k}_4(\bten{e}_\infty:\IB{r}\IB{r})(\IB{V}_r\cdot\IB{r})
		\right.
		\\
		&\qquad
		\left.+\Tilde{k}_5(\bten{e}_\infty\cdot\IB{r})\cdot(\bten{e}_\infty\cdot\IB{r})
		+ \Tilde{k}_6(\bten{e}_\infty:\IB{r}\IB{r})^2
		+ \Tilde{k}_7(\bten{e}_\infty\cdot\IB{r})\cdot(\bten{\omega}_\infty\cdot\IB{r})
		+ \Tilde{k}_8\bten{e}_\infty:\bten{e}_\infty
		\right],
		\nonumber
	\end{align}
	\begin{align}
		\IB{v}^{(1Q)}
		&= 4(1+\delta)\Big[
		\Tilde{j}_1\,(\bten{e}_{\infty}\!\cdot\!\IB{V}_r)
		+\Tilde{j}_2\,(\bten{e}_{\infty}\!\cdot\!\IB{r})(\IB{V}_r\!\cdot\!\IB{r})
		+ \Tilde{j}_3\,\IB{r}(\bten{e}_{\infty}\!:\!\IB{r}\IB{V}_r)
		+ \Tilde{j}_4\,\IB{V}_r(\bten{e}_{\infty}\!:\!\IB{r}\IB{r})
		\nonumber
		\\
		&\quad
		\nonumber
		+ \Tilde{j}_5\,\IB{r}(\IB{V}_r\!\cdot\!\IB{r})(\bten{e}_{\infty}\!:\!\IB{r}\IB{r})
		+ \Tilde{j}_6\,\bten{e}_{\infty}\!\cdot(\bten{e}_{\infty}\!\cdot\!\IB{r})
		+ \Tilde{j}_7\,\IB{r}(\bten{e}_{\infty}\!\cdot\!\IB{r})^{2}
		+ \Tilde{j}_8\,(\bten{e}_{\infty}\!\cdot\!\IB{r})(\bten{e}_{\infty}\!:\!\IB{r}\IB{r})
		\\
		&\quad
		+ \Tilde{j}_9\,\IB{r}(\bten{e}_{\infty}\!:\!\IB{r}\IB{r})^{2}
		+ \Tilde{j}_{10}\,\IB{r}(\bten{e}_{\infty}\!:\!\bten{e}_{\infty})
		\Big],
	\end{align}
	where the coefficients are provided in Supplementary material.

	\subsection{Force and torque corrections}
	From the flow fields derived at the first order, we evaluate the force and torque using:
	\begin{equation}
		\IB{F}^{(1)} = \int_{S_p} \IB{n} \cdot (-p^{(1)}\bten{I} + \bten{e}^{(1)} +\bten{\Pi}^{(0)}) \; dS
	\end{equation}
	\begin{equation} \label{eq: Torque 1}
		\IB{L}^{(1)} = \int_{S_p} \IB{r} \times (\IB{n} \cdot (-p^{(1)}\bten{I} + \bten{e}^{(1)} +\bten{\Pi}^{(0)})) \; dS
	\end{equation}
	On integrating the traction on the particle surface we get the first order force for buoyant and electrophoretic case as
	\begin{align}
		\IB{F}^{(1)}_B
		&= -6\pi  (1+\delta)\,\bten{e}_\infty \cdot\IB{V}_r
		+ 6\pi \delta(\bten{\omega}_\infty - \bten{e}_\infty)\cdot\IB{V}_r ,
		\label{eq:result-peery-B}
		\\
		\IB{F}^{(1)}_{E}
		&= -3\pi (1+\delta)\,\bten{e}_\infty \cdot\IB{V}_r . \label{eq:result-peery-E}
	\end{align}
	We found no torque corrections in either case at $O(\mathrm{Wi})$.
	Although Eq.~(\ref{eq: Torque 1}) produces a non-trivial intermediate expression, it vanishes identically because it involves the double contraction of the symmetric tensor and an antisymmetric Levi-Civita tensor ($\mathcal{L}_\epsilon$),
	$
	\frac{8}{21} \pi (1+\delta) \mathcal{L}_\epsilon\IB{:}(\bten{e_\infty}\cdot\bten{e_\infty}).
	$

	\subsection{Lift and drag corrections from the reciprocal theorem}
	
	To ascertain our calculations from earlier section, we also employ the reciprocal theorem to evaluate the force and torque corrections at $O(\mathrm{Wi})$.
	For this, we introduce two distinct test fields, $(p^t,\IB{v}^{t})$, respectively. These obey steady Stokes equations with the following boundary conditions:
	\begin{align}
		\IB{v}^{t} \big|_{r=1} &=
		\begin{cases}
			\widehat{\IB{V}}, & \text{(for extracting force correction)}, \\
			\widehat{\bten{\omega}} \times \IB{r}, & \text{(for extracting torque correction)}.
		\end{cases}\label{Testfield}
	\end{align}
	Following the earlier approaches of applying the reciprocal theorem to the first-order inhomogeneous Stokes equations \citep{KimKarrila13}, we obtain (\ms{detailed in Appendix C}):
	\begin{equation}\label{LRT:combined}
		\widehat{\IB{V}} \cdot \IB{F}^{(1)}
		=
		- \int_{V_f} \bten{\Pi}^{(0)} : \nabla \IB{V}^{t}  dV, 
		\qquad 
		\widehat{\bten{\omega}} \cdot \IB{L}^{(1)}
		=
		- \int_{V_f} \bten{\Pi}^{(0)} : \nabla \IB{V}^{t}  dV.
	\end{equation}
    On considering three independent orientations of both the test fields and the externally-induced $\IB{V}_r$, the resulting resistance tensor for buoyancy and electrophoresis takes the following structure for the coordinate setup in Fig.\ref{fig:schematic}: 
	\begin{align}
		\IB{F}^{(1)}_B = \bten{R}_B \cdot \IB{V}_r \mathcal,  \quad  \text{ where } \IB{r}_{B} &=
		\left[
		\begin{array}{ccc}
			0 & 0 & -3\pi\,\beta\,(1+\delta) \\
			0 & 0 & 0 \\
			-3\pi\,\beta\,(1+3\delta) & 0 & 0
		\end{array}
		\right],
		\label{eq:result-lrt-B}
		\\
		\IB{F}^{(1)}_E = \bten{R}_E \cdot \IB{V}_r \mathcal,  \quad  \text{ where }      \IB{r}_{E} &=
		\left[
		\begin{array}{ccc}
			0 & 0 & -\frac{3\pi}{2}\,\beta\,(1+\delta) \\
			0 & 0 & 0 \\
			-\frac{3\pi}{2}\,\beta \,(1+\delta) & 0 & 0
		\end{array}
		\right],
		\label{eq:result-lrt-E}
	\end{align}
	where $\mathsf{R}_{ij}$ is the $O(\text{Wi})$ force correction in $i^{th}$ direction arising from a field applied along the $j^{th}$ direction.
	Hence, $\mathsf{R}_{31}$ yields the lift ($z$-direction) correction for imposed relative velocity along streamwise ($x$) direction, whereas $\mathsf{R}_{13}$ denotes the drag ($x$) correction for relative velocity imposed along the velocity gradient ($z$) direction.
	Similarly, we constructed the rotational resistance tensor and found no $O(\mathrm{Wi})$ corrections, irrespective of the orientations of the test field and applied forcing mechanisms. These results match exactly with Eq.(\ref{eq:result-peery-B},\ref{eq:result-peery-E}) for the flow field setup in Fig.\ref{fig:schematic}.

\ms{
A striking feature of Eq. (\ref{eq:result-lrt-B},\ref{eq:result-lrt-E}) is that the electrophoretic resistance tensor ($\bten{R}_E$) is symmetric, whereas the buoyancy-driven tensor ($\bten{R}_B$) is not. Physically, the symmetry of $\bten{R}_E$ implies a reciprocal coupling: the cross-stream lift generated by a unit flow-directed translation (relative to the shear) equals the flow-direction drag generated by a unit cross-stream translation. 
This symmetry can be understood by decomposing the first-order viscoelastic force correction into contributions from the quadratic and co-rotational components. The quadratic stress preserves the symmetry of the extensional response. In contrast, the co-rotational stress accounts for the rotation of the polymer configuration with the fluid vorticity \cite{bird1986dynamics}; it is this kinematic coupling with the hydrodynamic signature of the tuning mechanism that generates the asymmetry in $\bten{R}$.
The hydrodynamic singularities of the two tuning mechanisms dictate which stress components survive. 
For electrophoresis, the far-field disturbance is characterized by an irrotational source dipole with no pressure disturbance \citep{morrison1970electrophoresis}. The absence of local vorticity causes the co-rotational component to vanish entirely ($\IB{F}^{(1C)}_E = \IB{0}$), leaving only the symmetric quadratic contribution. 
Conversely, the buoyancy-driven tuning generates the stokeslet that exhibits local vorticity. This contributes to the co-rotational stress $(6\pi \delta(\bten{\omega}_\infty - \bten{e}_\infty)\cdot\IB{V}_r)$, which generates the asymmetry in $\bten{R}_B$. The quadratic stress contributes $-6\pi   (1+\delta)\,\bten{e}_\infty \cdot\IB{V}_r$.
These findings suggest that when the hydrodynamic signature of the tuning mechanism is irrotational, we observe symmetry in  O(\text{Wi}) lift-drag corrections.
}


	\section{Discussion and Concluding Remarks}

	The analytical results presented in preceding section establish that different tuning mechanisms give rise to distinct lift and drag contributions. Here, we focus on understanding the leading-order lift force generated by tuning the particle to lead in streamwise direction.
	Given that $\delta$ usually lies between $-0.6$ and $-0.5$ \citep{bird1986dynamics}, Eqs.~(\ref{eq:result-peery-B},\ref{eq:result-lrt-B}) indicate that a particle 
	driven by the buoyant mechanism will undergo positive cross-stream migration: toward regions of higher velocity \ms{(towards $+\IB{e}_z$)}.
	In the limit $\delta = -0.5$, our result agrees exactly with that of 
	\citet{einarsson2017spherical} for an Oldroyd-B fluid, in which the second normal stress coefficient vanishes\footnote[2]{It should be noted that the relaxation times of the 
		second-order and Oldroyd-B fluid models are related by 
		$\lambda = 2\mu_r \lambda_{eff}$, where $\mu_r$ is the ratio of the polymer to the total solution viscosity.}.
	On the other hand, Eq.(\ref{eq:result-peery-E},\ref{eq:result-lrt-E}) predict that an electrophoretic mechanism would yield migration towards lower shear velocity \ms{(towards $-\IB{e}_z$)}.
	

		\begin{figure}
		\centering
		\includegraphics[width=1\linewidth]{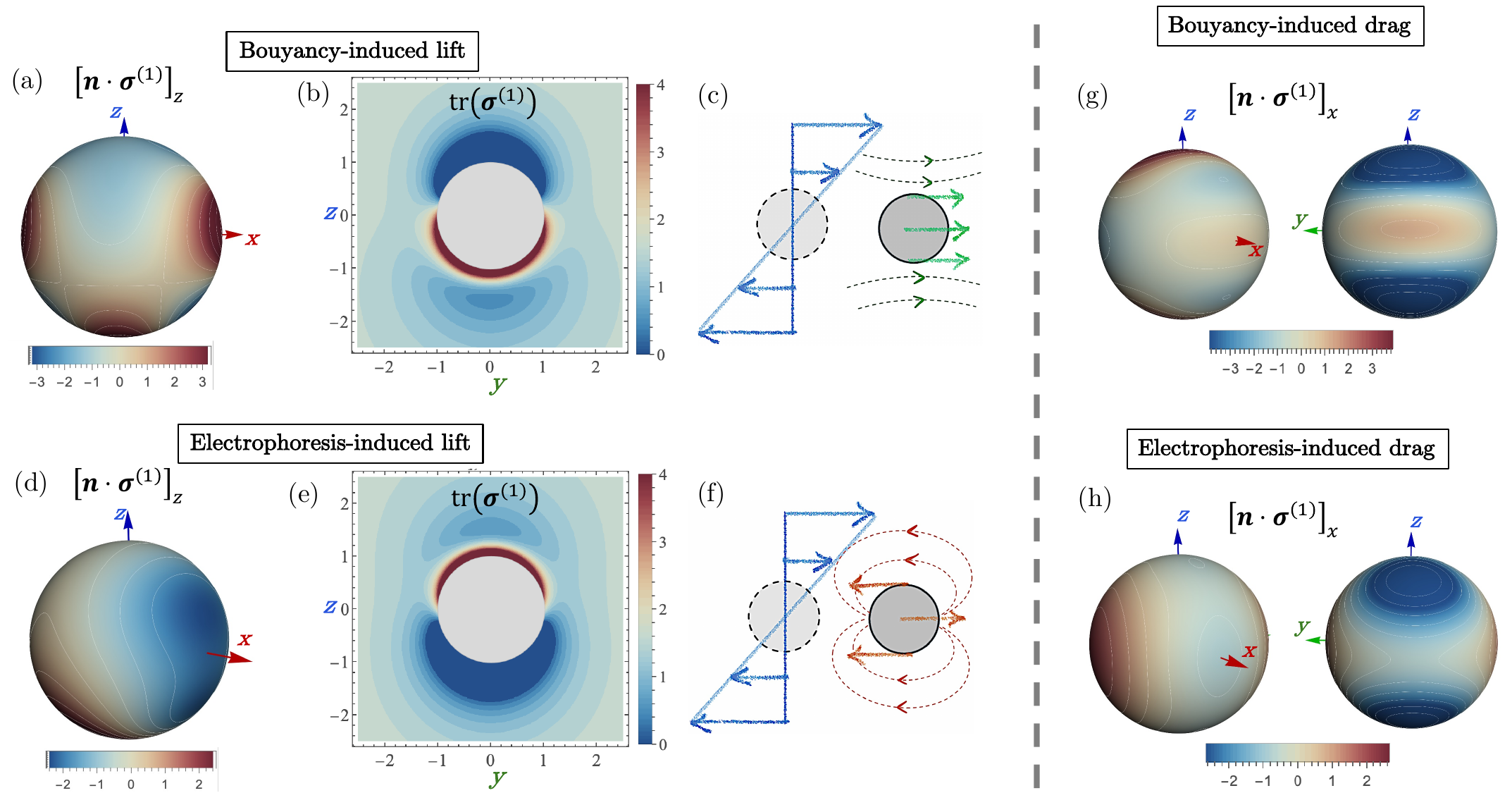}
		\caption{
			(a,d) Traction distribution of the lift ($z-$direction force correction) experienced by a particle made to lead the background shear along $x$-direction with $\mathcal{B} = 1 $ and $ \mathcal{E} = 1$, respectively.
			(b,e) Trace components of $O$(Wi) stress indicating the anisotropy in polymeric stretch in vorticity-gradient ($y-z$) plane.
			(c,f) Schematics illustrating the leading-order disturbance fields of force-bearing and force-free mechanisms (stokeslet and source dipole, respectively), and their induced relative velocity across the particle surface. The source dipole decays substantially more rapidly than the stokeslet; schematics are not drawn to scale.
			(g,h) Traction distribution of the drag correction ($x-$direction force correction) for a particle moving relative to the background shear along $z$-direction with $\mathcal{B} = 1 $ and $ \mathcal{E} = 1$, respectively. \ms{Complete viewpoints of the traction plots can be found in Appendix D.}
		} 
			\label{fig:result}
	\end{figure}

	Fig. \ref{fig:result}(a,d) shows $z-$component of traction ($\IB{n} \cdot \bten{\sigma}^{(1)}$, where $\bten{\sigma}^{(1)} = p^{(1)}\bten{I} + \bten{e}^{(1)} + \bten{\Pi}^{(0)}$) for both tuning mechanisms, highlighting the spatial distribution of lift forces across the particle surface.
	Next, Fig. \ref{fig:result}(b,e) compares the profile for trace of polymeric stress, which indicates the extent of polymer stretching around the particle. 
	As suggested by the qualitative arguments of \citet{Zhang2020}, for a leading buoyant particle, the increased relative velocity produces higher polymeric tension on the lower surface (Fig. \ref{fig:result}(b)), generating a lift force in $z-$direction.
	This argument, however, does not readily apply to a leading electrophoretic particle because the local disturbance velocity induced by the particle plays a crucial role, as evidenced by the traction distributions shown in Fig. \ref{fig:result}(d).
\ms{Qualitative illustration in Fig.2(c) depicts a particle leading the flow via buoyancy: due to the no-slip condition, the entire surface moves with the leading velocity (solid green arrows) relative to shear. Consequently, the particle experiences greater shear with respect to background flow on its bottom surface than on its top surface.
In contrast, Fig. 2(f) depicts an electrophoretic particle leading the flow: owing to the backward slip condition (up to $-3/2$ times the particle velocity \citep{masliyah2006electrokinetic}), the particle experiences greater shear on its top surface than on its bottom surface. 
These inversions of relative shear near the surface yield opposite trends of polymer stretching.
}
    \ms{Furthermore, as noted earlier, the disturbance field of a buoyant particle is fundamentally different in the vicinity of particle as compared to that generated by electrophoresis (as shown in Fig.1 and dashed arrows in Fig.2(c,f)). These reasons} limit the scope of a direct mechanistic analogy between the two scenarios.
	Hence in viscoelastic flows, the hydrodynamic signature of a \textit{driven} particle dictates the polymeric stress distribution and, consequently, the direction of the force corrections.
	Additionally, Fig. \ref{fig:result}(g,h) shows the traction plots for of cross-stream application of external field. We find that the drag corrections for both the forced and force-free mechanisms act in the same direction:
	a particle moving from low to high velocity regions (under the influence of external field) will experience a backward $x-$direction drag correction; a particle tuned to move from high- to low-velocity regions will experience a forward drag.


	The theoretical predictions presented here are corroborated by recent 
	experimental observations of electrokinetic particle manipulation in 
	microfluidic flows of PEO solution \citep{Serhatlioglu2020, Li2023}, 
	conducted at low Weissenberg numbers and polymer concentrations below 
	the overlap concentration.
	However, one may argue that at dilute polymer concentration, perhaps inertial lift \citep{kim2009Three,cevheri2014,choudhary2019inertial,KhairKabarowski2020}, which acts in the same direction, is focusing the particles. 
	To quantify this, we estimate the channel length that would be required by inertial and viscoelastic lifts to focus the particles in the setup of \citet{Serhatlioglu2020}. 
    \ms{Before presenting the order-of-magnitude argument, we clarify that our uniform shear flow results are compared with experiments that have Poiseuille flow profiles. This is justified for two reasons: (i) earlier studies have shown that flow curvature does not contribute to electrokinetic lift \citep{choudhary2019inertial,Choudhary20}, and (ii) the particle radius is an order of magnitude smaller than the length scale of the shear non-uniformity, ensuring the particle locally experiences a uniform shear.}
	The inertial lift velocity is predicted to be $\approx 0.29\,{\varepsilon|\zeta|a^{2}\rho\dot{\gamma}E^{\infty}}/{\mu^{2}}$, 
	yielding a lift velocity of $0.4\ \mu$m/s for $E^\infty = 200\ \text{V/cm}$, 
	$\varepsilon = 6.95\times10^{-10}\ \text{F/m}$, $|\zeta| = 80\ \text{mV}$, 
	$a = 3\ \mu\text{m}$, $\rho = 1000\ \text{kg/m}^3$, $\overline{\dot{\gamma}} = 140\ \text{s}^{-1}$, 
	and $\mu = 10^{-3}\ \text{Pa.s}$. 
	The axial particle velocity reported in their experiments is $5.2\ \text{mm/s}$, 
	which gives a focusing length of $\approx 38\ \text{cm}$; this is much larger than  observation window ($4\ \text{cm}$) in their experiments. 
	Next, we estimate the focusing length for viscoelastic lift for 100~ppm PEO ($c/c^* = 0.21$, $\lambda_{eff} = 8.9\ \text{ms}$, $\mu_r=0.033$). 
	In dimensional form, Eq.(\ref{eq:result-peery-E}) for typical value of $\delta=-0.5$ yields the lift velocity as $0.25\,\mu_r \dot{\gamma}^2 a \lambda_{eff}$. Substituting the values from experiments gives a velocity of $4.3 \mu$m/s, which yields focusing length to be within the observation window of their experiments $\approx$ 3.6 cm.
	Hence, in the weakly nonlinear regime, the observed focusing enhancement 
	is likely dominated by viscoelastic effects.
	\ms{Additionally, these findings indicate that the physical mechanisms driving the opposing migration directions reported by \citet{Li2018, Ma2026} likely stem from higher-order effects in Weissenberg number and polymer concentration, that further can trigger significant modifications to the surface slip and electrophoretic mobility within polymeric fluids \citep{LiKoch2020,ghosh2021electrophoretic,zhai2026electrokinetic}.}
	
	Beyond microfluidic particle manipulation, these results have broader implications 
	for understanding the locomotion of microswimmers in viscoelastic environments, 
	such as biological fluids. A self-propelled `neutral' squirmer \citep{lauga2020fluid}, representing \textit{Paramecium}, moves in a force-free manner with far-field source-dipole signature, making it a biological analogue of an electrophoretic particle. 
	The key result of this work, that force-free and force-bearing mechanisms can generate qualitatively distinct polymeric stresses in sheared flows, thus implies that such swimmers will potentially experience viscoelastic force and torque corrections. Moreover, since `pushers' and `pullers' carry their own distinct hydrodynamic  signatures, each swimmer type will yield its own characteristic viscoelastic modifications, offering a richer picture of navigation and transport in complex biological fluids.
	
	
	\vspace{1cm}
	
	\noindent
	\textbf{Acknowledgements}: The authors thank the Indian Institute of Technology Kanpur for support via Initiation Grant (IITK-CHE-2023066).

	\appendix
\renewcommand{\theequation}{A.\arabic{equation}}
\setcounter{equation}{0}

\newpage
\small

\ms{
\section*{Appendix A. Corotational and quadratic stress components}
\noindent
The actual or total flow field is governed by the following dimensionless equations (with the non-dimensionalization as prescribed in the main article): $  \nabla^2 \IB{V} - \nabla P = -Wi\, \nabla \cdot \bten{S}, \;
\nabla \cdot \IB{V} = 0,$
where $\IB{V}$, $P$, and $\bten{S}$ denote the actual flow velocity, pressure, and polymeric stress tensor, respectively.
Following Ganesh and Koch~\cite{KOCH200687}, the stress tensor $\bten{S}$ can be decomposed as: $\bten{S} = \bten{S}^{Q} + \bten{S}^{C} $,
where $\bten{S}^{Q}$ and $\bten{S}^{C}$ represent the quadratic and co-rotational contributions to the second-order fluid stress. For steady incompressible flow, these are defined as 
\begin{equation}
\bten{S}^{Q}
= 4(1+\delta)\,\bten{E}\cdot\bten{E},
\qquad \qquad
\bten{S}^{C}
= 2\delta
\left(
\nabla\cdot\left(\IB{V}\,\bten{E}\right)
+ \bten{W}\cdot\bten{E}
+ \bten{E}\cdot\bten{W}^{\mathsf{T}}
\right).
\label{Eq:actGanesh actual stress}
\end{equation}
Here, $\bten{E}$ and $\bten{W}$ are the strain and the rotation rate tensors for the actual flow field.
Next we split the actual flow into undisturbed flow and disturbance components as $\IB{V} = \IB{V}_{\infty} + \IB{v},
\;
P = P_{\infty} + p,
\; 
\bten{E} = \bten{e}_{\infty} + \bten{e},
\;
\bten{W} = \bten{\omega}_{\infty} + \bten{\omega}.$
Substituting these in Eq.(\ref{Eq:actGanesh actual stress}) we obtain
\begin{equation}
\begin{aligned}
\bten{S}^{Q}
&=
4(1+\delta)
\Big(
\bten{e}_{\infty}\cdot\bten{e}_{\infty}
+\bten{e}_{\infty}\cdot\bten{e}
+\bten{e}\cdot\bten{e}_{\infty}
+\bten{e}\cdot\bten{e}
\Big),
\\[6pt]
\bten{S}^{C}
&=
2\delta
\Big[\nabla\cdot(\IB{V}_{\infty}\bten{e}_{\infty})
+
\nabla\cdot(\IB{V}_{\infty}\bten{e})
+
\nabla\cdot(\IB{v}\bten{e}_{\infty})
+
\nabla\cdot(\IB{v}\bten{e})
+
\bten{\omega}_{\infty}\cdot\bten{e}_{\infty}
+
\bten{\omega}_{\infty}\cdot\bten{e}
+
\bten{\omega}\cdot\bten{e}_{\infty}
+
\bten{\omega}\cdot\bten{e}
\\
&\qquad
+
\bten{e}_{\infty}\cdot\bten{\omega}_{\infty}^{\mathsf T}
+
\bten{e}_{\infty}\cdot\bten{\omega}^{\mathsf T}
+
\bten{e}\cdot\bten{\omega}_{\infty}^{\mathsf T}
+
\bten{e}\cdot\bten{\omega}^{\mathsf T}
\Big].
\end{aligned}
\label{eq:SqSc_expanded}
\end{equation}
The undisturbed components are given by
\begin{equation}
\begin{aligned}
\bten{S}_{\infty}^{Q}
&=
4(1+\delta)\,
\bten{e}_{\infty}\cdot\bten{e}_{\infty},
\qquad
\qquad
\bten{S}_{\infty}^{C}
&=
2\delta
\Big[
\nabla\cdot
\left(
\IB{V}_{\infty}\bten{e}_{\infty}
\right)
+
\bten{\omega}_{\infty}\cdot\bten{e}_{\infty}
+
\bten{e}_{\infty}\cdot\bten{\omega}_{\infty}^{\mathsf T}
\Big],
\end{aligned}
\label{eq:Sinf}
\end{equation}
which yield no contribution to the flow profile as they are divergence free \cite{ho1976migration}.
The disturbance contributions to the non-Newtonian stress due to presence of particle are given by
\begin{equation}
\begin{aligned}
\bten{\Pi}^{Q}
&=
4(1+\delta)
\Big(
\bten{e}_{\infty}\cdot\bten{e}
+\bten{e}\cdot\bten{e}_{\infty}
+\bten{e}\cdot\bten{e}
\Big),
\\[6pt]
\bten{\Pi}^{C}
&=
2\delta
\Big[
\nabla\cdot(\IB{V}_{\infty}\bten{e})
+
\nabla\cdot(\IB{v}\bten{e}_{\infty})
+
\nabla\cdot(\IB{v}\bten{e})
+
\bten{\omega}_{\infty}\cdot\bten{e}
+
\bten{\omega}\cdot\bten{e}_{\infty}
+
\bten{\omega}\cdot\bten{e}
+
\bten{e}_{\infty}\cdot\bten{\omega}^{\mathsf T}
+
\bten{e}\cdot\bten{\omega}_{\infty}^{\mathsf T}
+
\bten{e}\cdot\bten{\omega}^{\mathsf T}
\Big].
\end{aligned}
\label{eq:PiQC}
\end{equation}
}

\ms{\section*{Appendix B: Buoyant particle in simple shear}}
Here we detail the approach mentioned in \S III. To calculate the first order lift and drag contributions. 
We first find out the right hand side of Eq.(\ref{eq:p1Q-rhs}) with known cofficients $f_i(r)$. \footnote[3]{We perform these evaluations in Mathematica and our framework makes use of commands such as `partialr', `laplacer', `contract', `prettyprint', `getNewIndex', `getScalarPrefactors' and `unitsphereIntegral' from open source `Matte' package developed by \citet{einarsson2017computer}.}

\begin{align}
	f_1(r) &= \frac{9}{8}\left(\frac{30}{r^{10}}-\frac{6}{r^{8}}+\frac{1}{r^{6}}\right), \notag\hspace{3pt} &
	f_2(r) &= \frac{9}{4}\left(\frac{25}{r^{12}}-\frac{36}{r^{10}}+\frac{6}{r^{8}}\right), \notag\\[3pt]
	f_3(r) &= -\left(\frac{450}{r^{12}}+\frac{234}{r^{10}}+\frac{105}{4r^{8}}-\frac{9}{r^{5}}\right), \notag\hspace{3pt} &
	f_4(r) &= \frac{15}{2}\left(\frac{60}{r^{14}}-\frac{84}{r^{12}}-\frac{16}{r^{10}}+\frac{3}{r^{7}}\right), \notag\\[3pt]
	f_5(r) &= 25\left(\frac{54}{r^{14}}-\frac{48}{r^{12}}+\frac{9}{r^{10}}+\frac{4}{r^{7}}\right), \notag\hspace{3pt} &
	f_6(r) &= -\frac{25}{2}\left(\frac{63}{r^{16}}+\frac{72}{r^{14}}-\frac{15}{r^{12}}+\frac{14}{r^{9}}\right), \notag\\[3pt]
	f_7(r) &= \left(\frac{75}{r^{12}}-\frac{40}{r^{10}}+\frac{25}{4r^{8}}-\frac{10}{r^{5}}\right).\notag
\end{align}
Based on the structure of these seven non-homogeneous terms, we form the ansatz for $p^{(1Q)}$ with unknown coefficients $g_i(r)$. 
\begin{align}
	&{p}^{(1Q)}
	=
	4(1+\delta)\left[
	g_1\IB{V}_r\cdot\IB{V}_r
	+ g_2(\IB{r}\cdot\IB{V}_r)^2
	+ g_3(\bten{e}_\infty:\IB{r}\IB{V}_r)
	+ g_4(\bten{e}_\infty:\IB{r}\IB{r})(\IB{r}\cdot\IB{V}_r)
	\nonumber
	\right.
	\\
	&\qquad
	\left.+ g_5(\bten{e}_\infty\cdot\IB{r})\cdot(\bten{e}_\infty\cdot\IB{r})
	+ g_6(\bten{e}_\infty:\IB{r}\IB{r})^2
	+ g_7\bten{e}_\infty:\bten{e}_\infty
	\right],
	\nonumber
\end{align}
This ansatz is then substituted in Eq.(\ref{eq:p1Q-GE}), which yields seven second-order ordinary differential equations in $g_i(r)$. 
{
	\begin{align}
		2g_2+\frac{2g_1'}{r}+g_1''
		&=\frac{9}{8}\left(\frac{30}{r^{10}}-\frac{6}{r^{8}}+\frac{1}{r^{6}}\right), \quad
		\frac{10g_6'}{r}+g_6''=-\frac{25}{2}\left(-\frac{63}{r^{16}}+\frac{72}{r^{14}}-\frac{15}{r^{12}}+\frac{14}{r^{9}}\right),
		\nonumber\\
		4g_4+\frac{4g_3'}{r}+g_3''
		&=-\frac{450}{r^{12}}+\frac{234}{r^{10}}-\frac{105}{4r^{8}}-\frac{9}{r^{5}}, \quad 
		\frac{8g_4'}{r}+g_4''=\frac{15}{2}\left(-\frac{60}{r^{14}}+\frac{84}{r^{12}}-\frac{16}{r^{10}}+\frac{3}{r^{7}}\right), 
		\nonumber\\
		8g_6+\frac{6g_5'}{r}+g_5''
		&=25\left(\frac{54}{r^{14}}-\frac{48}{r^{12}}+\frac{9}{r^{10}}+\frac{4}{r^{7}}\right), \quad 
		\frac{6g_2'}{r}+g_2''=\frac{9}{4}\left(\frac{25}{r^{12}}-\frac{36}{r^{10}}+\frac{6}{r^{8}}\right) ,
		\nonumber \\
		2g_7+\frac{2g_7'}{r}+g_7''
		&=\left(\frac{75}{r^{12}}-\frac{40}{r^{10}}+\frac{25}{4r^{8}}-\frac{10}{r^{5}}\right).
		\nonumber
	\end{align}
}
These equations can be readily solved to obtain the particular and homogeneous solutions, such that we substitute $g_i(r) = g_i^{h}(r) + g_i^{p}(r)$ (superscripts denote the homogeneous and particular components, respectively).

{
	\begin{align}
		g_1^{p} &= \frac{9}{32}\left(\frac{2}{r^{8}}-\frac{1}{r^{4}}\right),
		&\qquad
		g_1^{h} &= -\frac{C_{f1}}{3r^{3}}-\frac{C_{f2}}{r}, \notag
		\\
		g_2^{p} &= \frac{9}{8}\left(\frac{1}{r^{10}}-\frac{3}{r^{8}}+\frac{2}{r^{6}}\right),
		&
		g_2^{h} &= \frac{C_{f1}}{r^{5}}, \notag
		\\
		g_3^{p} &= \frac{3}{40}\left(-\frac{80}{r^{10}}+\frac{50}{r^{8}}+\frac{25}{r^{6}}+\frac{12}{r^{3}}\right),
		&
		g_3^{h} &= -\frac{C_{f3}}{5r^{5}}-\frac{C_{f4}}{3r^{3}}, \notag
		\\
		g_4^{p} &= -\frac{3}{4}\left(\frac{10}{r^{12}}-\frac{28}{r^{10}}+\frac{20}{r^{8}}+\frac{3}{r^{5}}\right),
		&
		g_4^{h} &= \frac{C_{f3}}{2r^{7}}, \notag
		\\
		g_5^{p} &= \left(\frac{15}{r^{12}}-\frac{20}{r^{10}}+\frac{25}{8r^{8}}\right),
		&
		g_5^{h} &= \frac{4C_{f5}}{63r^{7}}-\frac{C_{f6}}{5r^{5}}, \notag
		\\
		g_6^{p} &= \frac{5}{4}\left(\frac{9}{r^{14}}-\frac{20}{r^{12}}+\frac{15}{r^{10}}+\frac{10}{r^{7}}\right),
		&
		g_6^{h} &= -\frac{C_{f5}}{9r^{9}}, \notag
		\\
		g_7^{p} &= \left(\frac{1}{2r^{10}}-\frac{5}{3r^{3}}\right),
		&
		g_7^{h} &= -\frac{2C_{f5}}{315r^{5}}+\frac{C_{f6}}{15r^{3}}-\frac{C_{f7}}{r}. \notag
	\end{align}
}
Here $C_{fi}$ are the integration constants which will be later collectively found by using appropriate boundary conditions.
With the quadratic pressure field, substitution into Eq.~\eqref{eq:quad GE} yields the ansatz for the ${O}(\mathrm{Wi})$ quadratic velocity field $\IB{v}^{(1Q)}$. The velocity is expanded with \ms{coefficients $h_i(r)$ (comprising of contributions from $\nabla p^{(1Q)}-\nabla\cdot\bten{\Pi}^{(0Q)}$)}
\begin{align}
	= \Big[
	& h_1\IB{r}(\IB{V}_r\!\cdot\!\IB{V}_r)
	+ h_2\IB{V}_r(\IB{V}_r\!\cdot\!\IB{r})
	+ h_3\IB{r}(\IB{V}_r\!\cdot\!\IB{r})^{2}
	+ h_4(\bten{e}_{\infty}\!\cdot\!\IB{V}_r)
	+ h_5(\bten{e}_{\infty}\!\cdot\!\IB{r})(\IB{V}_r\!\cdot\!\IB{r})
	\nonumber
	\\
	&
	+ h_6\IB{V}_r(\bten{e}_{\infty}\!:\!\IB{r}\IB{r})
	+ h_7\IB{r}(\bten{e}_{\infty}\!:\!\IB{r}\IB{V}_r)
	+ h_8\IB{r}(\IB{V}_r\!\cdot\!\IB{r})(\bten{e}_{\infty}\!:\!\IB{r}\IB{r})
	+ h_9\bten{e}_{\infty}\!\cdot(\bten{e}_{\infty}\!\cdot\!\IB{r})
 \nonumber	\\
	&
	+ h_{10}\,\IB{r}(\bten{e}_{\infty}\!\cdot\!\IB{r})^{2}
	+ h_{11}(\bten{e}_{\infty}\!\cdot\!\IB{r})(\bten{e}_{\infty}\!:\!\IB{r}\IB{r})
	+ h_{12}\IB{r}(\bten{e}_{\infty}\!:\!\IB{r}\IB{r})^{2}
	+ h_{13}\IB{r}(\bten{e}_{\infty}\!:\!\bten{e}_{\infty})
	\Big]4(1+\delta). \nonumber
\end{align}
\ms{where $h_i(r)$ coefficients detailed as follows:}
{
	\begin{align}
		h_1(r) &= -\frac{9}{16r^{8}} + \frac{9}{8r^{6}} + \frac{C_{f1}}{r^{5}} + \frac{C_{f2}}{r^{3}}, \notag\hspace{3pt} &
		h_2(r) &= \frac{9}{16}\!\left(-\frac{5}{r^{8}} + \frac{6}{r^{6}}\right)
		+ \frac{2C_{f1}}{r^{5}}, \notag\\[3pt]
		h_3(r) &= \frac{9}{4}\!\left(\frac{2}{r^{10}} - \frac{3}{r^{8}}\right)
		- \frac{5C_{f1}}{r^{7}}, \notag\hspace{3pt} &
		h_4(r) &= \frac{27}{8r^{8}} + \frac{15}{8r^{6}} + \frac{3}{2r^{5}}
		- \frac{3}{5r^{3}} - \frac{C_{f3}}{5r^{5}} - \frac{C_{f4}}{3r^{3}}, \notag\\[3pt]
		h_5(r) &= \frac{1}{4}\!\left(\frac{36}{r^{10}} - \frac{75}{r^{8}} - \frac{30}{r^{7}} + \frac{9}{r^{5}}
		+ \frac{4C_{f3}}{r^{7}}\right), \notag\hspace{3pt} &
		h_6(r) &= \frac{1}{4}\!\left(\frac{54}{r^{10}} - \frac{45}{r^{8}} - \frac{15}{r^{7}}
		- \frac{9}{r^{5}} + \frac{2C_{f3}}{r^{7}}\right), \notag\\[3pt]
		h_7(r) &= \frac{1}{8}\!\left(\frac{12}{r^{10}} - \frac{105}{r^{8}}
		- \frac{60}{r^{7}} +\frac{8C_{f3}}{r^{7}} - \frac{9}{20r^{5}}
		+ \frac{C_{f4}}{r^{5}}\right), \notag\hspace{3pt} &
		h_8(r) &= -\frac{75}{2r^{12}} + \frac{60}{r^{10}}
		+ \frac{105}{4r^{9}} - \frac{7C_{f3}}{2r^{9}}, \notag\\[3pt]
		h_9(r) &=-\frac{35}{2r^{10}} - \frac{20}{r^{7}} + \frac{15}{r^{5}}
		+ \frac{8C_{f5}}{63r^{7}} - \frac{2C_{f6}}{5r^{5}}, \notag\hspace{3pt} &
		h_{10}(r) &=\frac{25}{2r^{12}} + \frac{25}{r^{10}}
		+ \frac{70}{r^{9}} - \frac{4C_{f5}}{9r^{9}} - \frac{75}{2r^{7}} + \frac{C_{f6}}{r^{7}}, \notag\\[3pt]
		h_{11}(r) &= \frac{25}{4}\!\left(-\frac{7}{r^{12}} + \frac{10}{r^{10}}\right)
		+ \frac{70}{r^{9}} - \frac{4C_{f5}}{9r^{9}}, \notag\hspace{3pt} &
		h_{12}(r) &= \frac{75}{r^{14}} - \frac{125}{r^{12}}
		- \frac{315}{2r^{11}} + \frac{C_{f5}}{r^{11}}, \notag\\[3pt]
		h_{13}(r) &=-\frac{5}{2r^{10}} - \frac{5}{r^{7}} + \frac{15}{2r^{5}}
		+ \frac{2C_{f5}}{63r^{7}}
		- \frac{C_{f6}}{5r^{5}} + \frac{C_{f7}}{r^{3}}.
		\nonumber
	\end{align}
}
Following the similar procedure used to determine the pressure field, we now take the Laplacian of the $\IB{v}^{(1Q)}$ ansatz, which is made using the unknown coefficients $j_i(r)$. This results in a system of thirteen second-order ordinary differential equations detailed below.
{
	\begin{align}
		2j_3 + \frac{4}{r}j_1' + j_1'' 
		&= -\frac{9}{16r^{8}} + \frac{9}{8r^{6}} + \frac{C_{f1}}{r^{5}} + \frac{C_{f2}}{r^{3}},\nonumber \hspace{15pt}
		4j_3 + \frac{4}{r}j_2' + j_2'' = \frac{9}{16}\!\left(-\frac{5}{r^{8}} + \frac{6}{r^{6}}\right)
		+ \frac{2C_{f1}}{r^{5}},\nonumber
		\\[3pt]
		\frac{8}{r}j_3' + j_3'' 
		&= \frac{9}{4}\!\left(\frac{2}{r^{10}} - \frac{3}{r^{8}}\right)
		- \frac{5C_{f1}}{r^{7}},\nonumber \hspace{15pt}
		4j_8 + \frac{6}{r}j_5' + j_5'' = \frac{1}{4}\!\left(\frac{36}{r^{10}} - \frac{75}{r^{8}} - \frac{30}{r^{7}} + \frac{9}{r^{5}}
		+ \frac{4C_{f3}}{r^{7}}\right),\nonumber
		\\[3pt]
		2j_8 + \frac{6}{r}j_6' + j_6'' 
		&= \frac{1}{4}\!\left(\frac{54}{r^{10}} - \frac{45}{r^{8}} - \frac{15}{r^{7}}
		- \frac{9}{r^{5}} + \frac{2C_{f3}}{r^{7}}\right),\nonumber \hspace{15pt}
		\frac{10}{r}j_8' + j_8'' = -\frac{75}{2r^{12}} + \frac{60}{r^{10}}
		+ \frac{105}{4r^{9}} - \frac{7C_{f3}}{2r^{9}},\nonumber
		\\[3pt]
		8j_{12} + \frac{8}{r}j_{11}' + j_{11}'' 
		&= \frac{25}{4}\!\left(-\frac{7}{r^{12}} + \frac{10}{r^{10}}\right)
		+ \frac{70}{r^{9}} - \frac{4C_{f5}}{9r^{9}},\nonumber \hspace{15pt}
		\frac{12}{r}j_{12}' + j_{12}'' = \frac{75}{r^{14}} - \frac{125}{r^{12}}
		- \frac{315}{2r^{11}} + \frac{C_{f5}}{r^{11}},\nonumber
		\\[3pt]
		2j_5 + 2j_7 + \frac{2}{r}j_4' + j_4'' 
		&= \frac{27}{8r^{8}} + \frac{15}{8r^{6}} + \frac{3}{2r^{5}}
		- \frac{3}{5r^{3}} - \frac{C_{f3}}{5r^{5}} - \frac{C_{f4}}{3r^{3}},\nonumber
		\\[3pt]
		4j_8 + \frac{6}{r}j_7' + j_7'' 
		&= \frac{1}{8}\!\left(\frac{12}{r^{10}} - \frac{105}{r^{8}}
		- \frac{60}{r^{7}} +\frac{8C_{f3}}{r^{7}} - \frac{9}{20r^{5}}
		+ \frac{C_{f4}}{r^{5}}\right),\nonumber
		\\[3pt]
		4j_{10} + 4j_{11} + \frac{4}{r}j_{9}' + j_{9}'' 
		&= -\frac{35}{2r^{10}} - \frac{20}{r^{7}} + \frac{15}{r^{5}}
		+ \frac{8C_{f5}}{63r^{7}} - \frac{2C_{f6}}{5r^{5}},\nonumber
		\\[3pt]
		8j_{12} + \frac{8}{r}j_{10}' + j_{10}'' 
		&= \frac{25}{2r^{12}} + \frac{25}{r^{10}}
		+ \frac{70}{r^{9}} - \frac{4C_{f5}}{9r^{9}} - \frac{75}{2r^{7}} + \frac{C_{f6}}{r^{7}},\nonumber
		\\[3pt]
		2j_{10} + \frac{4}{r}j_{13}' + j_{13}'' 
		&= -\frac{5}{2r^{10}} - \frac{5}{r^{7}} + \frac{15}{2r^{5}}
		+ \frac{2C_{f5}}{63r^{7}}
		- \frac{C_{f6}}{5r^{5}} + \frac{C_{f7}}{r^{3}}.
		\nonumber
	\end{align}
}
On solving we can next write $ \IB{v}^{(1Q)}$ in terms of $j_i(r)$.
We get the quadratic component of first order velocity field as:
\begin{equation}
	\begin{aligned}
		\IB{v}^{(1Q)}
		&= 4(1+\delta)\Big[
		j_1\,\IB{r}(\IB{V}_r\!\cdot\!\IB{V}_r)
		+ j_2\,\IB{V}_r(\IB{V}_r\!\cdot\!\IB{r})
		+ j_3\,\IB{r}(\IB{V}_r\!\cdot\!\IB{r})^{2}
		+ j_4\,(\bten{e}_{\infty}\!\cdot\!\IB{V}_r)
		\\
		&\quad
		+ j_5\,(\bten{e}_{\infty}\!\cdot\!\IB{r})(\IB{V}_r\!\cdot\!\IB{r})
		+ j_6\,\IB{V}_r(\bten{e}_{\infty}\!:\!\IB{r}\IB{r})
		+ j_7\,\IB{r}(\bten{e}_{\infty}\!:\!\IB{r}\IB{V}_r)
		+ j_8\,\IB{r}(\IB{V}_r\!\cdot\!\IB{r})(\bten{e}_{\infty}\!:\!\IB{r}\IB{r})
		\\
		&\quad
		+ j_9\,\bten{e}_{\infty}\!\cdot(\bten{e}_{\infty}\!\cdot\!\IB{r})
		+ j_{10}\,\IB{r}(\bten{e}_{\infty}\!\cdot\!\IB{r})^{2}
		+ j_{11}\,(\bten{e}_{\infty}\!\cdot\!\IB{r})(\bten{e}_{\infty}\!:\!\IB{r}\IB{r})
		+ j_{12}\,\IB{r}(\bten{e}_{\infty}\!:\!\IB{r}\IB{r})^{2}
		+ j_{13}\,\IB{r}(\bten{e}_{\infty}\!:\!\bten{e}_{\infty})
		\Big]. \nonumber
	\end{aligned}
\end{equation}
where  $j_i(r) = j_i^{h}(r) + j_i^{p}(r)$, and the coefficients $j_i(r)$ $(i=1,\ldots,13)$ are decomposed into contributions from the homogeneous and particular solutions. The explicit expressions for these coefficients are given below.
{
	\[
	\begin{aligned}
		j_1^{p} &= -\frac{3}{32}\!\left(\frac{1}{r^{6}}+\frac{3}{r^{4}}\right),
		&\qquad
		j_1^{h} &= -\frac{C_{f2}}{2r}+\frac{C_{h1}}{35r^{5}}-\frac{C_{h2}}{3r^{3}},
		\\[3pt]
		j_2^{p} &= -\frac{9}{32}\!\left(\frac{1}{r^{6}}+\frac{1}{r^{4}}\right),
		&
		j_2^{h} &= \frac{2C_{h1}}{35r^{5}}-\frac{C_{h3}}{3r^{3}},
		\\[3pt]
		j_3^{p} &= \frac{9}{16}\!\left(\frac{1}{r^{8}}+\frac{2}{r^{6}}\right),
		&
		j_3^{h} &= \frac{C_{f1}}{2r^{5}}-\frac{C_{h1}}{7r^{7}},
		\\[3pt]
		j_4^{p} &= \frac{1}{10r}-\frac{5}{8r^{4}}-\frac{3}{28r^{3}},
		&
		j_4^{h} &= \frac{C_{f3}}{70r^{3}}+\frac{C_{f4}}{18r}+\frac{C_{h4}}{r}
		-\frac{C_{h5}}{3r^{3}}+\frac{2C_{h8}}{35r^{5}},
		\\[3pt]
		j_5^{p} &= \frac{1}{r^{8}}+\frac{15}{8r^{6}}+\frac{15}{28r^{5}}-\frac{3}{8r^{3}},
		&
		j_5^{h} &= -\frac{C_{f3}}{14r^{5}}+\frac{C_{h5}}{r^{5}}
		-\frac{C_{h7}}{r^{5}}-\frac{2C_{h8}}{7r^{7}},
	\end{aligned}
	\\[3pt]\]
	\[
	\begin{aligned}
				j_6^{p} &= \frac{1}{56}\!\left(\frac{49}{r^{8}}+\frac{35}{r^{6}}+\frac{15}{r^{5}}+\frac{21}{r^{3}}\right),
		&
		j_6^{h} &= -\frac{C_{f3}}{28r^{5}}+\frac{C_{h6}}{r^{5}}
		-\frac{C_{h8}}{7r^{7}},
		\\[3pt]
		j_7^{p} &= \frac{11}{16r^{8}}+\frac{45}{16r^{6}}+\frac{15}{28r^{5}}+\frac{3}{40r^{3}},
		&
		j_7^{h} &= -\frac{C_{f3}}{14r^{5}}-\frac{C_{f4}}{6r^{3}}
		+\frac{C_{h7}}{r^{5}}-\frac{2C_{h8}}{7r^{7}},
		\\[3pt]
		j_8^{p} &= -\frac{15}{8}\!\left(\frac{2}{r^{10}}+\frac{4}{r^{8}}+\frac{1}{r^{7}}\right),
		&
		j_8^{h} &= \frac{C_{f3}}{4r^{7}}+\frac{C_{h8}}{r^{9}},
		\\[3pt]
		j_9^{p} &= \frac{25}{8r^{6}}-\frac{2}{r^{5}},
		&
		j_9^{h} &= \frac{4C_{f5}}{315r^{5}}-\frac{8C_{h9}}{693r^{7}} 
		+\frac{2C_{h10}}{35r^{5}}+\frac{2C_{h11}}{35r^{5}}
		-\frac{C_{h12}}{3r^{3}},
		\\[3pt]
		j_{10}^{p} &= \frac{5}{8}\!\left(-\frac{2}{r^{10}}-\frac{15}{r^{8}}+\frac{6}{r^{5}}\right),
		&
		j_{10}^{h} &= -\frac{C_{f6}}{10r^{5}}+\frac{4C_{h9}}{99r^{9}}
		-\frac{C_{h10}}{7r^{7}},
		\\[3pt]
		j_{11}^{p} &= -\frac{25}{16}\!\left(\frac{2}{r^{10}}+\frac{3}{r^{8}}\right),
		&
		j_{11}^{h} &= \frac{4C_{h9}}{99r^{9}}-\frac{C_{h11}}{7r^{7}},
		\\[3pt]
		j_{12}^{p} &= \frac{5}{4}\!\left(5\!\left(\frac{1}{r^{12}}+\frac{2}{r^{10}}\right)+\frac{7}{r^{9}}\right),
		&
		j_{12}^{h} &= -\frac{C_{f5}}{18r^{9}}-\frac{C_{h9}}{11r^{11}},
		\\[3pt]
		j_{13}^{p} &= \frac{25}{24r^{6}}-\frac{1}{2r^{5}},
		&
		j_{13}^{h} &= \frac{C_{f5}}{315r^{5}}-\frac{C_{f7}}{2r}
		-\frac{C_{h10}}{693r^{7}}+\frac{C_{h13}}{35r^{5}}-\frac{C_{h13}}{3r^{3}}.
		\\[3pt]
	\end{aligned}
	\]
}
On applying the boundary conditions mentioned in Eq. (\ref{eq:model_eq_first}) and continuity equation, we evaluate the coefficients $C_{fi}$ and $C_{hi}$, respectively; finally merging the particular and homogeneous solution we get the complete expressions for $g_i(r)$ and $j_i(r)$ for both $p^{(1Q)}$ and $\IB{v}^{(1Q)}$.
{
	\begin{align}
		g_{1} &= \left(\frac{9}{16r^8}-\frac{9}{32r^4}+\frac{3}{16r^3}\right) \notag 
		&\quad
		g_{2} &= \left(-\frac{9}{8r^{10}}+\frac{27}{8r^8}-\frac{9}{4r^6}+\frac{9}{16r^5}\right) \notag \\
		g_{3} &= \left(-\frac{6}{r^{10}}+\frac{15}{4r^8}+\frac{15}{8r^6}-\frac{49}{16r^5}+\frac{3}{8r^3}\right) \notag &\quad
		g_{4} &= -\frac{15}{2 r^{12}}+\frac{21}{r^{10}}-\frac{15}{r^8}+\frac{245}{32 r^7}-\frac{9}{4 r^5} \notag \\
		g_{5} &= \left(\frac{15}{r^{12}}-\frac{20}{r^{10}}+\frac{25}{8r^8}+\frac{55}{4r^7}-\frac{86}{r^5}\right) \notag &\quad
		g_{6} &= \left(\frac{45}{4r^{14}}-\frac{25}{r^{12}}+\frac{75}{4r^{10}}-\frac{385}{16r^9}+\frac{25}{2r^7}\right) \notag \\
		g_{7} &= \left(\frac{1}{2r^{10}}-\frac{11}{8r^5}+\frac{25}{28r^3}\right)\notag\\
		j_{1} &= \left(-\frac{3}{32 r^6}+\frac{9}{32 r^5}-\frac{9}{32 r^4}+\frac{3}{32 r^3}\right) \notag &\quad
		j_{2} &= \left(\frac{9}{32 r^6}-\frac{9}{16 r^5}+\frac{9}{32 r^4}\right) \notag \\
		j_{3} &= \left(-\frac{9}{16 r^8}+\frac{45}{32 r^7}-\frac{9}{8 r^6}+\frac{9}{32 r^5}\right) \notag &\quad
		j_{4} &= \left(\frac{17}{32 r^5}-\frac{5}{8 r^4}-\frac{3}{32 r^3}+\frac{3}{16 r}\right) \notag \\
		j_{5} &= \left(\frac{1}{r^8}-\frac{85}{32 r^7}+\frac{15}{8 r^6}+\frac{5}{32 r^5}-\frac{3}{8 r^3}\right) \notag &\quad
		j_{6} &= \left(\frac{7}{8 r^8}-\frac{85}{64 r^7}+\frac{5}{8 r^6}-\frac{35}{64 r^5}+\frac{3}{8 r^3}\right)\notag\\
		j_{7} &= \left(\frac{11}{16 r^8}-\frac{85}{32 r^7}+\frac{45}{16 r^6}-\frac{21}{32 r^5}-\frac{3}{16 r^3}\right)\notag&\quad
		j_{8} &= \left(\frac{15}{4 r^{10}}+\frac{595}{64 r^9}-\frac{15}{2 r^8}+\frac{125}{64 r^7}\right) \notag \\
		j_{9} &= \left(\frac{55}{28 r^7}-\frac{25}{8 r^6}+\frac{65}{56 r^5}\right) \notag &\quad
		j_{10} &= \left(\frac{5}{4 r^{10}}-\frac{55}{8 r^9}+\frac{75}{8 r^8}-\frac{215}{56 r^7}+\frac{5}{56 r^5}\right) \notag \\
		j_{11} &= \left(-\frac{25}{8 r^{10}}+\frac{55}{8 r^9}-\frac{75}{16 r^8}+\frac{15}{16 r^7}\right) \notag &\quad
		j_{12} &= \left(-\frac{25}{4 r^{12}}+\frac{495}{32 r^{11}}-\frac{25}{2 r^{10}}+\frac{105}{32 r^9}\right) \notag \\
		j_{13} &= \left(-\frac{55}{112 r^7}+\frac{25}{24 r^6}-\frac{65}{112 r^5}+\frac{5}{168 r^3}  \right)\notag
	\end{align}
}

The coefficients for the corotational component $k_i(r)$, where $k_i(r)$ $(i=1,\ldots,8)$ are given as:
\small{
	\begin{align}
		k_{1} &= \left(\frac{9}{4 r^{10}}-\frac{45}{8 r^8}-\frac{9}{4 r^6}\right)  \notag 
		\hspace{3pt} &
		k_{2} &= \left(-\frac{12}{r^{10}}+\frac{7}{2 r^8}-\frac{15}{2 r^6}-\frac{3}{r^5}+\frac{3}{2 r^3}\right) \notag \\[3pt]
		k_{3} &= \left(-\frac{15}{r^{12}}+\frac{34}{r^{10}}+\frac{15}{r^8}+\frac{15}{2 r^7}-\frac{9}{r^5}\right) \notag \hspace{3pt} &
		k_{4} &= \left(\frac{30}{r^{12}}-\frac{30}{r^{10}}+\frac{25}{2 r^8}+\frac{20}{r^7}-\frac{20}{r^5}\right) \notag \\[3pt]
		k_{5} &= \left(\frac{45}{2 r^{14}}-\frac{75}{2 r^{12}}-\frac{25}{2 r^{10}}-\frac{35}{r^9}+\frac{50}{r^7}\right) \notag \hspace{3pt} &
		k_{6} &= \left(-\frac{3}{2r^{3}}\right)
		\hspace{5pt}
		k_{7} =\left(\frac{10}{r^{5}}\right)
		\hspace{5pt}
		k_{8} = \left(\frac{1}{r^{10}}-\frac{2}{r^{5}}\right)
	\end{align}
} 
\\

\ms{
\section*{Appendix C. Details of the reciprocal theorem derivation}
\noindent
The test fields $(\text{velocity }\IB{v}^t, \text{ pressure }p^t, \text{ stress }\bten{\tau}^t)$ obey the Stokes equations and are subjected to the boundary condition described as Eq.(\ref{Testfield}). 
Following \citet{ho1976migration}, we proceed with the ${O}$(Wi) equations and subject it to an inner product with the test flow field. Similarly, we take an inner product of test field equations with the first order flow field.
\begin{equation}
\begin{aligned}
(\nabla \cdot \bten{\tau}^{(1)}) \cdot \IB{v}^{t}
&= - (\nabla \cdot \bten{\Pi}^{(0)}) \cdot \IB{v}^{t},
\qquad 
(\nabla \cdot \bten{\tau}^{t}) \cdot \IB{v}^{(1)}
= 0.
\end{aligned}
\label{eq:reciprocal_intermediate}
\end{equation} where,
    \begin{equation}
   \bten{\tau}^{(1)}= -p^{(1)} \bten{I} + \nabla \IB{v}^{(1)} + (\nabla \IB{v}^{(1)})^\mathsf{T},
   \qquad
    \bten{\tau}^{t}= -p^{t} \bten{I} + \nabla \IB{v}^{t} + (\nabla \IB{v}^{t})^\mathsf{T}
     \label{eq:Newtonian_stress}
    \end{equation}
 Employing the index notation, we first write each left-hand side in Eq.(\ref{eq:reciprocal_intermediate}) as
\begin{equation}
  \begin{aligned}
(\nabla\cdot{\bten{\tau}}^{(1)})\cdot\IB{v}^t
&=
(\partial_i\bten{\tau}^{(1)}_{ij})\IB{v}_j^t,
\qquad \qquad
(\nabla\cdot{\bten{\tau}}^{t})\cdot \IB{v}^{(1)}
&=
(\partial_i\bten{\tau}^{t}_{ij})\IB{v}_j^{(1)} ,
\end{aligned}  
\end{equation}
and similarly the right-hand side can be expressed. We next subtract the two equations of Eq. (\ref{eq:reciprocal_intermediate}):
\begin{equation}\label{eq:merged_lrt_app}
(\partial_i\bten{\tau}^{(1)}_{ij})\IB{v}_j^t
-
(\partial_i\bten{\tau}^{t}_{ij})\IB{v}_j^{(1)}
=
- (\partial_i \bten{\Pi}^{(0)}_{ij})  \IB{v}^{t}_j
\end{equation}
To simplify the left-hand side, we use the product rule to write: $(\partial_i\bten{\tau}^{(1)}_{ij})\IB{v}_j^t
=
\partial_i(\bten{\tau}^{(1)}_{ij}\IB{v}_j^t)-
\bten{\tau}^{(1)}_{ij}\partial_i \IB{v}_j^t
$, and 
$
(\partial_i\bten{\tau}^{t}_{ij})\IB{v}_j^{(1)}
=
\partial_i(\bten{\tau}^{t}_{ij}\IB{v}_j^{(1)})
-
\bten{\tau}^{t}_{ij}\partial_i \IB{v}_j^{(1)}
.
$
Therefore, Eq. (\ref{eq:merged_lrt_app}) becomes
\begin{equation}\label{eq:merged2_lrt_app}
    \partial_i(\bten{\tau}^{(1)}_{ij}\IB{v}_j^t)-
\bten{\tau}^{(1)}_{ij}\partial_i \IB{v}_j^t
-
\partial_i(\bten{\tau}^{t}_{ij}\IB{v}_j^{(1)})+
\bten{\tau}^{t}_{ij}\partial_i \IB{v}_j^{(1)}
=
- (\partial_i \bten{\Pi}^{(0)}_{ij})  \IB{v}^{t}_j
\end{equation}
Next, we focus on simplifying the second and fourth term on the left-hand side. For this, we employ the stress relation defined by Eq.(\ref{eq:Newtonian_stress}) and incompressibility condition to obtain
\begin{equation}
   \begin{aligned}
\bten{\tau}^{(1)}_{ij}\partial_i \IB{v}_j^t
=
(\partial_i \IB{v}_j^{(1)})(\partial_i \IB{v}_j^t)
+
(\partial_j \IB{v}_i^{(1)})(\partial_i \IB{v}_j^t),
\qquad
\bten{\tau}^{t}_{ij}\partial_i \IB{v}_j^{(1)}
=
(\partial_i \IB{v}_j^t)(\partial_i \IB{v}_j^{(1)})
+
(\partial_j \IB{v}_i^t)(\partial_i \IB{v}_j^{(1)}).
\end{aligned} 
\end{equation}
Since the dummy indices can be interchanged \(i\leftrightarrow j\) we find that $\bten{\tau}^{(1)}_{ij}\partial_i \IB{v}_j^t
=
\bten{\tau}^{t}_{ij}\partial_i \IB{v}_j^{(1)}$. Therefore Eq. (\ref{eq:merged2_lrt_app}) can be written as:
\begin{equation}\label{eq:merged3_lrt_app}
    \partial_i(\bten{\tau}^{(1)}_{ij}\IB{v}_j^t)
-
\partial_i(\bten{\tau}^{t}_{ij}\IB{v}_j^{(1)})
=
\bten{\Pi}^0_{ij}\partial_i \IB{v}_j^t-\partial_i(\bten{\Pi}^0_{ij}\IB{v}_j^t)
 ,
\end{equation}
where we have use the product rule on the right-hand side.
We rearrange the terms and express the above equation in vector notation as
\begin{equation}
\nabla\cdot
\Big[
(\bten{\tau}^{(1)}+\bten{\Pi}^{(0)})\cdot\IB{v}^{t}
-
 \bten{\tau}^{t}\cdot\IB{v}^{(1)}
\Big]
=
\bten{\Pi}^0:\nabla\IB{v}^t 
\end{equation}
Integrating over the fluid volume \(V_f\) and applying the divergence theorem on left-hand side yields,
\begin{equation}
- \int_{S_p}
\IB{n}\cdot
\Big[
(\bten{\tau}^{(1)}+\bten{\Pi}^0)\cdot\IB{v}^t
-
\bten{\tau}^t\cdot\IB{v}^{(1)}
\Big]
\, dS
= 
\int_{V_f}
\bten{\Pi}^0:\nabla\IB{v}^t
\, dV ,
\end{equation}
where $\IB{n}$ is the unit normal vector pointing into the fluid (hence the negative sign). 
 Within the resistance formulation, the particle is not allowed to migrate in the cross-stream direction while the hydrodynamic force corrections are evaluated. Thus, the second term vanishes, and we obtain
\begin{equation}
\widehat{\IB{V}}\cdot
\int_{S_p}
\IB{n}\cdot
(\bten{\tau}^{(1)}+\bten{\Pi}^0)
\,dS
=
-
\int_{V_f}
\bten{\Pi}^0:\nabla{\IB{v}}^t
\,dV.
\end{equation}
The above relation yields Eq. (\ref{LRT:combined}), which is used to capture force in a single direction for a single test field along the same axis in which tuning is applied (here `tuning' refers to the external-field-imparted relative velocity).
To capture the three components of force correction for a particle tuned in x-direction, we separately employ the test field corresponding to motion along three directions. We perform the same procedure for y-direction and z-direction tuning. Thus, we obtain nine elements that comprise the $\bten{R}$ tensor. We conduct this procedure to obtain buoyancy tuning ($\bten{R}_B$) and electrophoretic tuning ($\bten{R}_E$).
}




\newpage

\ms{\section*{Appendix D. Spatial distribution of traction}}

\begin{figure}[h]
    \centering
    \includegraphics[width=0.9\linewidth]{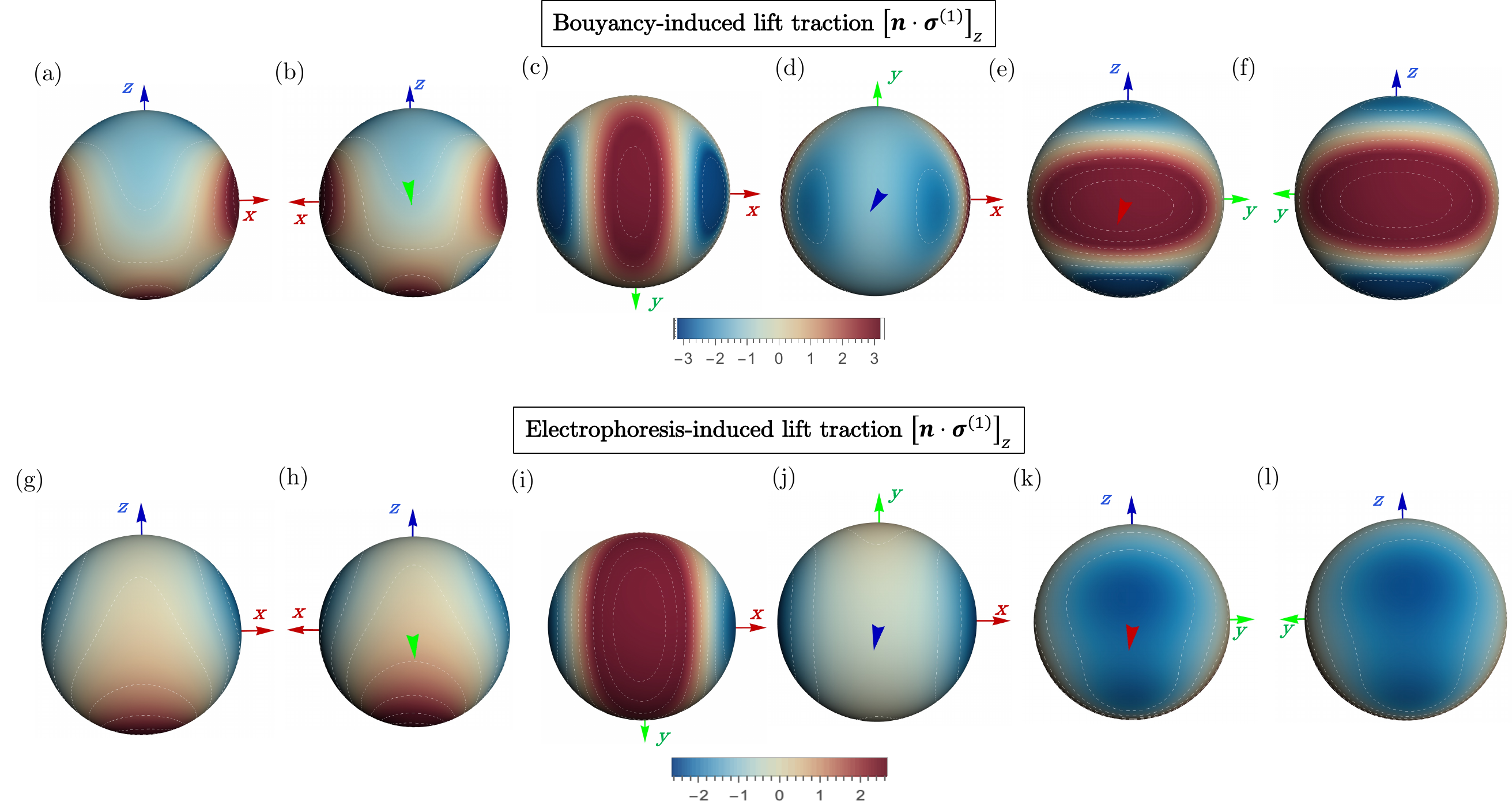}
    \includegraphics[width=0.9\linewidth]{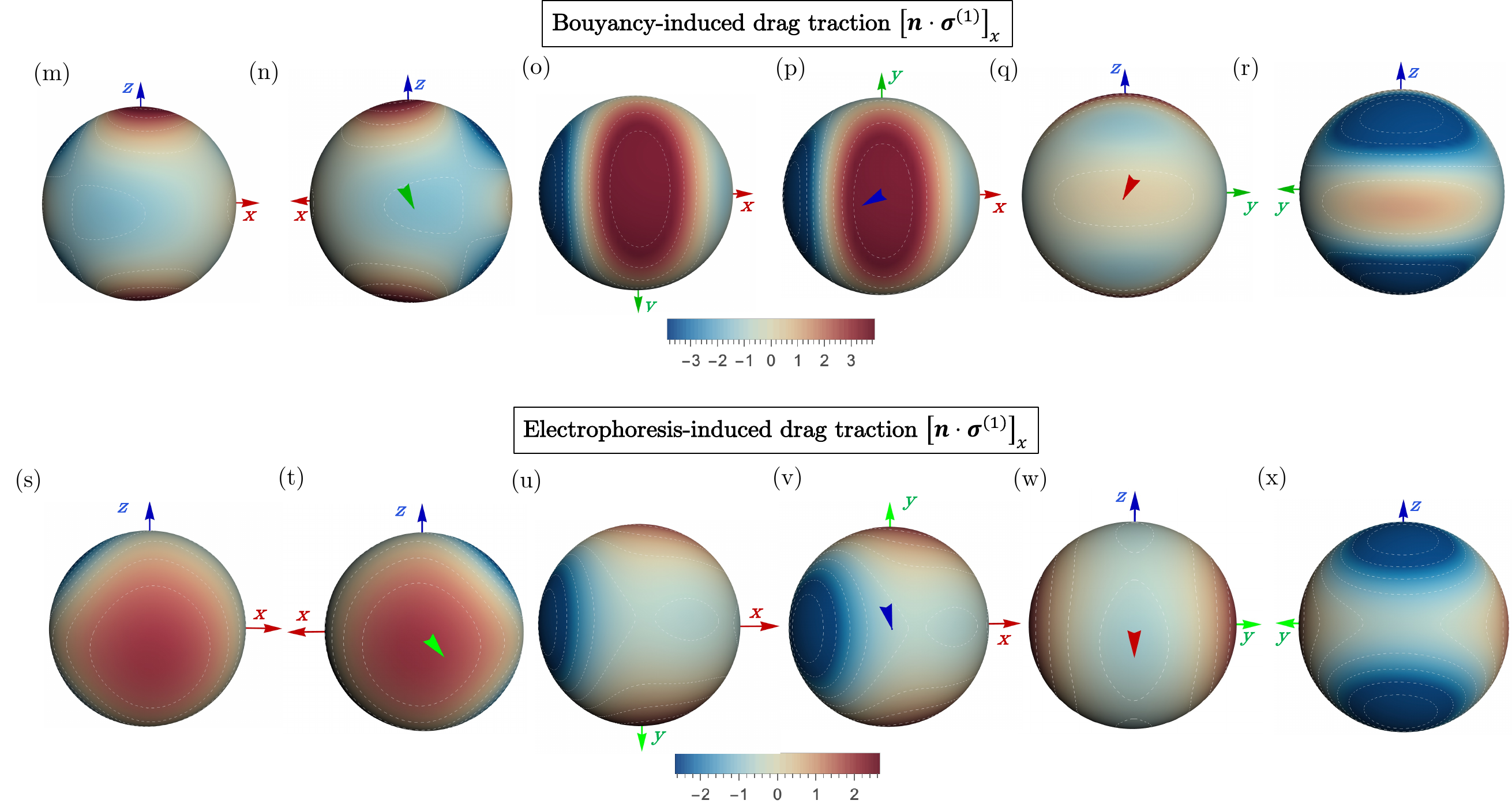}
    \caption{\footnotesize{
    \ms{(a-l) Spatial maps for visualization of distribution of z-traction $[\boldsymbol{n} \cdot \boldsymbol{\sigma}^{(1)}]_z$, that sums up to form the $\mathcal{O}(Wi)$ lift. 
    In addition to the analytical results presented in main text, the net lift force can be qualitatively assessed via summation: (a-f) show that the buoyancy-tuned particle exhibits a net surplus of positive (red) traction, yielding an overall positive lift, and (g-l) show the electrophoretic-tuned particle exhibits a predominantly negative (blue) traction field, resulting in a net negative lift.
    Similarly, spatial maps (m-r) and (s-x) show the distribution of x-traction. 
    }}
    }
    \label{fig:placeholder}
\end{figure}


\textbf{References}
\bibliography{apssamp}

\end{document}